\documentclass[useAMS,usenatbib]{mnras}
\usepackage{graphicx, bm, amssymb, xcolor}
\usepackage{verbatim}
\usepackage[all]{hypcap}
\usepackage{xspace}
\usepackage{multirow,bigdelim}
\usepackage[fleqn]{amsmath}
\usepackage{threeparttable}
\usepackage{epsfig}
\usepackage{aas_macros}
\usepackage{natbib}
\usepackage{times,txfonts}
\usepackage{morefloats}
\usepackage{tabularx}
\usepackage{lscape}
\usepackage{hyperref}
\usepackage{xcolor}
\usepackage{orcidlink}
\usepackage{adjustbox}
\usepackage{longtable}
\usepackage{deluxtable}
\usepackage{rotating}

\newcommand{\NII}{{\rm [N}$\,${\sc ii}{\rm ]}\xspace}
\newcommand{\OIII}{{\rm [O}$\,${\sc iii}{\rm ]}\xspace}
\newcommand{\coone}{\mbox{\rm CO($1\text{--}0$)}\xspace} 
\newcommand{\cotwo}{\mbox{\rm CO($2\text{--}1$)}\xspace} 

\newcommand{\Msun}{\mbox{M$_\odot$}}
\newcommand{\msun}{\mbox{M$_\odot$}}

\newcommand{\pyr}{\mbox{${\rm yr^{-1}}$}}

\newcommand{\pc}{\mbox{${\rm pc}$}}

\newcommand{\kms}{\mbox{${\rm km}~{\rm s}^{-1}$}}

\newcommand{\dex}{\mbox{${\rm dex}$}}

\newcommand{\vfb}{$v_{\rm fb}$}
\newcommand{\esf}{$\epsilon_{\rm sf}$}

\newcommand{\be}{\begin{equation}}
\newcommand{\ee}{\end{equation}}
\newcommand{\bea}{\begin{eqnarray}}
\newcommand{\eea}{\end{eqnarray}}

\newcommand{\appropto}{\mathrel{\vcenter{
  \offinterlineskip\halign{\hfil$##$\cr
    \propto\cr\noalign{\kern1pt}\sim\cr\noalign{\kern-2pt}}}}}

\usepackage{etoolbox}

\title[]{\vspace{-6mm}Environmental dependence of the molecular cloud lifecycle in 54 main sequence galaxies \vspace{-5mm}}

\author{Jaeyeon~Kim\orcidlink{0000-0002-0432-6847},$^1$\thanks{E-mail: \href{kim@uni-heidelberg.de}{kim@uni-heidelberg.de}}
M\'{e}lanie~Chevance\orcidlink{0000-0002-5635-5180},$^{1,2}$
J.~M.~Diederik~Kruijssen\orcidlink{0000-0002-8804-0212},$^1$
\newauthor
Adam K. Leroy,$^3$ Andreas~Schruba,$^4$ Ashley~T.~Barnes\orcidlink{0000-0003-0410-4504},$^{5}$ Frank Bigiel,$^5$ 
\newauthor
Guillermo A.~Blanc\orcidlink{0000-0003-4218-3944},$^{6,7}$  Yixian~Cao\orcidlink{0000-0001-5301-1326},$^4$
Enrico Congiu,$^7$ 
Daniel~A.~Dale\orcidlink{0000-0002-5782-9093},$^8$ 
\newauthor
Christopher M.~Faesi\orcidlink{0000-0001-5310-467X},$^{9}$ Simon~C.~O.~Glover\orcidlink{0000-0001-6708-1317},$^{2}$
Kathryn~Grasha\orcidlink{0000-0002-3247-5321},$^{10,11}$ Brent~Groves,$^{12}$
\newauthor
Annie Hughes,$^{13,14}$ Ralf~S.~Klessen\orcidlink{0000-0002-0560-3172},$^{2,15}$ Kathryn Kreckel\orcidlink{0000-0001-6551-3091},$^1$ Rebecca McElroy,$^{16}$
\newauthor
Hsi-An~Pan\orcidlink{0000-0002-1370-6964},$^{17}$ J\'{e}r\^{o}me Pety\orcidlink{0000-0003-3061-6546},$^{18,19}$ Miguel~Querejeta\orcidlink{0000-0002-0472-1011},$^{20}$
Alessandro~Razza\orcidlink{0000-0001-7876-1713},$^{7}$ 
\newauthor
Erik~Rosolowsky\orcidlink{0000-0002-5204-2259},$^{21}$
Toshiki~Saito\orcidlink{0000-0002-2501-9328},$^{22}$ Eva~Schinnerer\orcidlink{0000-0002-3933-7677},$^{23}$ Jiayi~Sun\orcidlink{0000-0003-0378-4667},$^{24,25}$
\newauthor
Neven Tomi\v{c}i\'{c},$^{26,27}$
Antonio Usero,$^{28}$
Thomas~G.~Williams\orcidlink{0000-0002-0012-2142}$^{23}$ 
\\\\
Affiliations are listed at the end of the paper
\vspace{-4mm}}
\begin{document}

\date{Accepted 2022 August 16. Received 2022 August 15; in original form 2022 June 14 \vspace{-3mm}}

\pagerange{\pageref{firstpage}--\pageref{lastpage}} \pubyear{2022}

\maketitle

\label{firstpage}

\begin{abstract} 

The processes of star formation and feedback, regulating the cycle of matter between gas and stars on the scales of giant molecular clouds (GMCs; $\sim$100\,pc), play a major role in governing galaxy evolution. Measuring the time-scales of GMC evolution is important to identify and characterise the specific physical mechanisms that drive this transition. By applying a robust statistical method to high-resolution CO and narrow-band H$\alpha$ imaging from the PHANGS survey, we systematically measure the evolutionary timeline from molecular clouds to exposed young stellar regions on GMC scales, across the discs of an unprecedented sample of 54 star-forming main-sequence galaxies (excluding their unresolved centres). We find that clouds live for about $1{-}3$ GMC turbulence crossing times ($5{-}30$~Myr) and are efficiently dispersed by stellar feedback within $1{-}5$~Myr once the star-forming region becomes partially exposed, resulting in integrated star formation efficiencies of $1{-}8$\%. These ranges reflect physical galaxy-to-galaxy variation. In order to evaluate whether galactic environment influences GMC evolution, we correlate our measurements with average properties of the GMCs and their local galactic environment. We find several strong correlations that can be physically understood, revealing a quantitative link between galactic-scale environmental properties and the small-scale GMC evolution. Notably, the measured CO-visible cloud lifetimes become shorter with decreasing galaxy mass, mostly due to the increasing presence of CO-dark molecular gas in such environment. Our results represent a first step towards a comprehensive picture of cloud assembly and dispersal, which requires further extension and refinement with tracers of the atomic gas, dust, and deeply-embedded stars.

\end{abstract}


\section{Introduction} \label{sec:intro}

Giant molecular clouds (GMCs) are the most important sites for star formation. The properties of the clouds are set by the large-scale environment of their host galaxies, directly linking the initial conditions of star formation to galactic-scale properties \citep{hughes13, colombo14, schruba19, sun18, sun20b, sun20}. In turn, the energy, momentum and metals deposited by stellar feedback drive the continuous evolution of the interstellar medium (ISM) in general \citep[e.g.][]{krumholz14}. The characterisation of the evolutionary time-scales from molecular cloud assembly to star formation, and to young stellar regions devoid of cold gas provides important insights into which physical mechanisms regulate this multi-scale cycle, and is therefore crucial to understanding the evolution of galaxies.

Theoretical studies of GMCs indicate that their evolution is influenced by various environmentally dependent dynamical properties such as gravitational collapse of the ISM, collisions between clouds, epicyclic motions, galactic shear, and large-scale gas streaming motions \citep[][]{dobbs13, dobbs15, meidt13,meidt18,meidt20, jeffreson18, jeffreson20, jeffreson21, tress20, tress21}. During recent decades, growing cloud-scale observational evidence, revealing a spatial decorrelation between molecular gas and young stellar regions, points towards a view of GMCs as transient objects that are dispersed within a free-fall or dynamical time-scale ($10-30$\,Myr) by violent feedback from young massive stars \citep{elmegreen00, engargiola03, blitz07, kawamura09, onodera10, schruba10, miura12, meidt15, corbelli17, kruijssen19, schinnerer19, chevance20_rev, chevance20,barnes20, kim21, pan22}. This contradicts a conventional view where GMCs are considered to represent quasi-equilibrium structures that survive over a large fraction of a galactic rotation period \citep{scoville79}. 

Despite these previous efforts, it has been challenging to understand what determines the evolutionary time-scales of cloud assembly, star formation, and cloud dispersal due to the limited range of galactic properties and ISM conditions probed so far. Now, the dynamic range of environments that can be investigated at required cloud-scale resolution has been significantly widened thanks to PHANGS\footnote{The Physics at High Angular resolution in Nearby GalaxieS project: http://phangs.org}, which has mapped $^{12}$CO($J{=}2{-}1)$ emission with the Atacama Large Millimeter/submillimeter Array (ALMA) at cloud-scale resolution across 90 star-forming main sequence galaxies \citep{leroy21_survey}. In addition, subsets of these galaxies have been targeted by observations at various other wavelengths including radio (PHANGS--VLA; A. Sardone et al. in prep.), mid-/near-infrared \citep{leroy19}, optical (PHANGS--HST; \citealp{lee22}, PHANGS--MUSE; \citealp{ emsellem22}, PHANGS--H$\alpha$; Razza et al. in prep.), and near-/far-ultraviolet \citep{leroy19}. These observations reveal that the GMC populations of nearby galaxies, with molecular gas surface densities spanning 3.4$\rm\,\dex$ \citep{sun20}, reside in diverse galactic environments covering a substantial range of local galaxy properties such as gas and stellar mass surface densities, orbital velocities and shear parameters  (J.~Sun et al., subm.).

\citet{onodera10} and \citet{schruba10} first quantified the spatial decorrelation of gas and young stellar regions observed at small scales in the Local Group galaxy M33. \citet{kruijssen14} and \citet{kruijssen18} developed a statistical method that translates the observed scale dependence of this spatial decorrelation between gas and young stellar regions into their underlying evolutionary timeline, ranging from cloud assembly to subsequent star formation and cloud dispersal, and finally to young stellar regions free of molecular gas. This method has utilised CO and H$\rm\alpha$ observations of nearby galaxies to characterise the evolutionary timeline between quiescent molecular gas to exposed young stellar regions for 15 galaxies \citep{kruijssen19, chevance20, zabel20, kim21, ward22}. So far, these measurements of time-scales, have been limited to a small number of galaxies due to the lack of CO imaging of star-forming discs at cloud-scale resolution and the fact that our method requires us to resolve at least the separation length between independent star-forming regions ($\rm 100{-}300\,pc$). These previous studies did not allow us to identify the key environmental factors and cloud properties (e.g. total, gas, molecular gas surface densities and masses) responsible for setting these time-scales. 

The first application of this method to a subset of nine PHANGS galaxies by \citet{chevance20} revealed that the time-scale for GMC survival is $10-30$\,Myr, agreeing well with the Local Group measurements by \citet{fukui08}, \citet{kawamura09}, and \citet{corbelli17}, using a different methodology. \citet{chevance20} also found that at high molecular gas surface density (with kpc-scale molecular gas surface density $\Sigma_{\rm H_{2}}>8\,M_{\odot}\,\rm pc^{-2}$), the measured cloud lifetime is consistent with being set by large-scale dynamical processes, such as large-scale gravitational collapse and galactic shear. In the low surface density regime ($\Sigma_{\rm H_{2}}\leq8\,M_{\odot}\,\rm pc^{-2}$), time-scales associated with internal dynamical processes, such as the free-fall and crossing times, govern the cloud lifetime. The duration over which CO and H$\rm\alpha$ emission overlap is found to be short ($1-5$\,Myr), indicating that pre-supernova feedback, such as photoionisation and stellar winds, plays a key role for molecular cloud disruption. This method also has been applied to other wavelengths. \citet{ward20_HI} used HI data to infer a duration of $\sim$50\,Myr for the atomic gas cloud lifetime in the LMC. For six nearby galaxies, \citet{kim21} incorporated \textit{Spitzer} 24$\rm\mu m$ observations to measure the time-scale of the heavily obscured star formation, which is missed when using H$\rm\alpha$ only, due to attenuation provided by surrounding gas and dust. The measured duration for the heavily obscured star formation is $1-4$\,Myr, constituting $10-25$\% of the cloud lifetime.

In this paper, we greatly increase the number of main sequence galaxies analysed by this statistical method from nine considered by \citet{chevance20} to 54 galaxies here. We capitalise on our CO observations from PHANGS--ALMA \citep{leroy21_survey} and a new, large narrow-band H$\rm\alpha$ survey by A. Razza et al. (in prep.; PHANGS--H$\alpha$). By applying our analysis to these galaxies, we systematically obtain the evolutionary sequence of GMCs from a quiescent molecular cloud phase to feedback dispersal phase, and finally to gas-free HII region phase. This statistically representative PHANGS sample covers a large range of galactic properties ($\sim$2\,dex in stellar mass) and morphologies. It enables us to quantitatively study the connection between the small-scale evolutionary cycle of molecular clouds and galactic-scale environmental properties. 

The structure of the paper is as follows. In Section~\ref{sec:data}, we summarise the observational data used in our analysis. In Section~\ref{sec:method}, we describe the statistical method used here and the associated main input parameters. In Section~\ref{sec:results}, we present the inferred cloud lifetime, the duration for which CO and H$\rm\alpha$ emission coincide, the mean separation length between star-forming regions undergoing independent evolution, as well as several other quantities derived from our measurements. In Section~\ref{sec:discussion}, we explore how these time-scales vary with galactic and average GMC properties. We also compare them with theoretical values. Lastly, we present our conclusions in Section~\ref{sec:concl}.

\section{Observational Data}\label{sec:data}

\begin{table*}
\begin{center}
\caption{Physical and observational properties of our galaxy sample.} \label{tab:sample}
\setlength\tabcolsep{9pt}
\begin{threeparttable}
\begin{adjustbox}{width=\textwidth}
\begin{tabular}{lccccccccc}
\hline
& (a) & (b) & (c) & (d) & (e) & (f) & (g) & (h)  & (i) \\
Galaxy & $M_{\rm *, global}$ &$\rm SFR_{global}$& $M_{\rm HI, global}$ & $M_{\rm H_{2}, global}$& $\rm\Delta$MS& Dist. & Incl. & P.A.  & Hubble  \\
&[$\rm log_{10}\,M_{\odot}$]& [$\rm log_{10}$\,$M_{\rm\odot}\rm yr^{-1}$]&[$\rm log_{10}\,M_{\odot}$]&[$\rm log_{10}\,M_{\odot}$]& & [Mpc] & [deg] &[deg]&type\\
\hline
IC1954&9.7&-0.4&8.8&8.7&-0.04&12.8&57.1&63.4&3.3\\
IC5273&9.7&-0.3&9.0&8.6&0.09&14.18&52.0&234.1&5.6\\
NGC0628&10.3&0.2&9.7&9.4&0.18&9.84&8.9&20.7&5.2\\
NGC0685&10.1&-0.4&9.6&8.8&-0.25&19.94&23.0&100.9&5.4\\
NGC1087&9.9&0.1&9.1&9.2&0.33&15.85&42.9&359.1&5.2\\
NGC1097&10.8&0.7&9.6&9.7&0.33&13.58&48.6&122.4&3.3\\
NGC1300&10.6&0.1&9.4&9.4&-0.18&18.99&31.8&278.0&4.0\\
NGC1365&11.0&1.2&9.9&10.3&0.72&19.57&55.4&201.1&3.2\\
NGC1385&10.0&0.3&9.2&9.2&0.50&17.22&44.0&181.3&5.9\\
NGC1433&10.9&0.1&9.4&9.3&-0.36&18.63&28.6&199.7&1.5\\
NGC1511&9.9&0.4&9.6&9.2&0.59&15.28&72.7&297.0&2.0\\
NGC1512&10.7&0.1&9.9&9.1&-0.21&18.83&42.5&261.9&1.2\\
NGC1546&10.4&-0.1&8.7&9.3&-0.15&17.69&70.3&147.8&-0.4\\
NGC1559&10.4&0.6&9.5&9.6&0.50&19.44&65.4&244.5&5.9\\
NGC1566&10.8&0.7&9.8&9.7&0.29&17.69&29.5&214.7&4.0\\
NGC1672&10.7&0.9&10.2&9.9&0.56&19.4&42.6&134.3&3.3\\
NGC1792&10.6&0.6&9.2&9.8&0.32&16.2&65.1&318.9&4.0\\
NGC1809&9.8&0.8&9.6&9.0&1.08&19.95&57.6&138.2&5.0\\
NGC2090&10.0&-0.4&9.4&8.7&-0.25&11.75&64.5&192.46&4.5\\
NGC2283&9.9&-0.3&9.7&8.6&-0.04&13.68&43.7&-4.1&5.9\\
NGC2835&10.0&0.1&9.5&8.8&0.26&12.22&41.3&1.0&5.0\\
NGC2997&10.7&0.6&9.9&9.8&0.31&14.06&33.0&108.1&5.1\\
NGC3059&10.4&0.4&9.7&9.4&0.29&20.23&29.4&-14.8&4.0\\
NGC3351&10.4&0.1&8.9&9.1&0.05&9.96&45.1&193.2&3.1\\
NGC3507&10.4&-0.0&9.3&9.3&-0.10&23.55&21.7&55.8&3.1\\
NGC3511&10.0&-0.1&9.4&9.0&0.06&13.94&75.1&256.8&5.1\\
NGC3596&9.7&-0.5&8.9&8.7&-0.12&11.3&25.1&78.4&5.2\\
NGC3627&10.8&0.6&9.1&9.8&0.19&11.32&57.3&173.1&3.1\\
NGC4254&10.4&0.5&9.5&9.9&0.37&13.1&34.4&68.1&5.2\\
NGC4298&10.0&-0.3&8.9&9.2&-0.18&14.92&59.2&313.9&5.1\\
NGC4303&10.5&0.7&9.7&9.9&0.54&16.99&23.5&312.4&4.0\\
NGC4321&10.7&0.6&9.4&9.9&0.21&15.21&38.5&156.2&4.0\\
NGC4496A&9.5&-0.2&9.2&8.6&0.28&14.86&53.8&51.1&7.4\\
NGC4535&10.5&0.3&9.6&9.6&0.14&15.77&44.7&179.7&5.0\\
NGC4540&9.8&-0.8&8.4&8.6&-0.46&15.76&28.7&12.8&6.2\\
NGC4548&10.7&-0.3&8.8&9.2&-0.58&16.22&38.3&138.0&3.1\\
NGC4569&10.8&0.1&8.8&9.7&-0.26&15.76&70.0&18.0&2.4\\
NGC4571&10.1&-0.5&8.7&8.9&-0.43&14.9&32.7&217.5&6.4\\
NGC4654&10.6&0.6&9.8&9.7&0.36&21.98&55.6&123.2&5.9\\
NGC4689&10.2&-0.4&8.5&9.1&-0.37&15.0&38.7&164.1&4.7\\
NGC4731&9.5&-0.2&9.4&8.6&0.30&13.28&64.0&255.4&5.9\\
NGC4781&9.6&-0.3&8.9&8.8&0.09&11.31&59.0&290.0&7.0\\
NGC4941&10.2&-0.4&8.5&8.7&-0.30&15.0&53.4&202.2&2.1\\
NGC4951&9.8&-0.5&9.2&8.6&-0.14&15.0&70.2&91.2&6.0\\
NGC5042&9.9&-0.2&9.3&8.8&0.01&16.78&49.4&190.6&5.0\\
NGC5068&9.4&-0.6&8.8&8.4&0.02&5.2&35.7&342.4&6.0\\
NGC5134&10.4&-0.3&8.9&8.8&-0.45&19.92&22.7&311.6&2.9\\
NGC5248&10.4&0.4&9.5&9.7&0.25&14.87&47.4&109.2&4.0\\
NGC5530&10.1&-0.5&9.1&8.9&-0.37&12.27&61.9&305.4&4.2\\
NGC5643&10.3&0.4&9.1&9.4&0.36&12.68&29.9&318.7&5.0\\
NGC6300&10.5&0.3&9.1&9.3&0.13&11.58&49.6&105.4&3.1\\
NGC6744&10.7&0.4&10.3&9.5&0.06&9.39&52.7&14.0&4.0\\
NGC7456&9.6&-0.4&9.3&9.3&-0.02&15.7&67.3&16.0&6.0\\
NGC7496&10.0&0.4&9.1&9.3&0.53&18.72&35.9&193.7&3.2\\
\hline
\end{tabular}
\end{adjustbox}
\begin{tablenotes}
\item (a) \& (b) Stellar mass and global SFR \citep{leroy21_survey}. (c) Atomic gas mass from Lyon-Meudon Extragalactic Database (LEDA). (d) Aperture corrected total molecular gas mass from PHANGS--ALMA observations \citep{leroy21_pipe}. (e) Offset from the star-forming main sequence \citep{leroy21_survey}. (f) Distance \citep{anand21}. (g) \& (h) Inclination and Position angle \citep{lang20}. (i)  Hubble type from LEDA.  
\end{tablenotes}
\end{threeparttable}
\end{center}
\end{table*}

\subsection{Descriptions of CO and H$\alpha$ emission maps}
PHANGS has constructed a multi-wavelength database at GMC-scale resolution ($\sim$100\,pc), covering most of the nearby ($\leq$20\,Mpc), ALMA accessible, star-forming galaxies ($M_{*}=10^{9.5}{-}10^{11.5} \,M_{\odot}$) lying around the main sequence (see \citealp{leroy21_survey}). In this paper, we focus on the galaxies where both $^{12}$CO($J{=}2{-}1$), denoted as \cotwo in the following, and ground-based continuum-subtracted H$\rm\alpha$ observations are available. This results in a sample of 64 galaxies. For a robust application of our statistical method, we need a minimum of 35 identified emission peaks in each map \citep{kruijssen18}. This requirement made us remove 10 galaxies (IC\,5332, NGC\,1317, NGC\,2566, NGC\,2775, NGC\,3626, NGC\,4207, NGC\,4293, NGC\,4424, NGC\,4457, NGC\,4694), as they do not have enough peaks identified in either the CO or H$\rm\alpha$ map\footnote{Most of these galaxies have centrally concentrated star formation making it hard to distinguish emission peaks. Excluding these galaxies does not bias our galaxy sample in terms of stellar mass, as they seem to be distributed evenly across the observed range (see Figure~\ref{fig:exsig}).}. In the end, our final sample consists of 54 galaxies and their physical and observational properties are listed in Table~\ref{tab:sample}. In the next paragraphs, we briefly summarise the main features of the PHANGS--ALMA (\cotwo) and PHANGS--H$\rm\alpha$ data sets.

In order to trace the molecular gas, we use the the PHANGS--ALMA survey, which has mapped the \cotwo emission in the star-forming part of the disc across 90 galaxies. Full descriptions of the sample and the survey design are presented in \citet{leroy21_survey}. Detailed information about the image production process can be found in \citet{leroy21_pipe}. The observations were carried out using 12-m, 7-m, and total power antennas of the ALMA. The resulting maps have a resolution of $\sim$1\arcsec, which translates into a physical scale of $\sim$25-200~pc for the galaxies considered here. We use the first public release version of moment-0 maps generated with an inclusive signal masking scheme to ensure a high detection completeness (the ``broad'' masking scheme; see \citealp{leroy21_pipe}) at native resolution. The typical 1$\sigma$ surface density sensitivity of these broad maps is  $\Sigma_{\rm H_{2}}\approx 5.8~\rm M_{\odot}\,\rm{pc^{-2}}$ (assuming the Galactic \coone-to-$\rm H_{2}$ conversion factor of 4.35\,$M_{\rm\odot}\,\rm pc^{-2}(K\,km\,s^{-1})^{-1}$ and a \cotwo-to-\coone ratio of 0.65; \citealp{leroy13, denbrok21, leroy21_lineratio}).

To trace the star formation rate (SFR), we use the continuum-subtracted narrow-band H$\rm\alpha$ imaging from PHANGS--H$\rm\alpha$ (Preliminary version; A. Razza et al. in prep.). We assume that all the H$\alpha$ emission originates from the gas ionised by young, massive stars, ignoring the contributions from other sources such as supernova remnants and planetary nebulae. For the 19 galaxies in PHANGS--H$\rm\alpha$ that were also surveyed by PHANGS--MUSE \citep{emsellem22}, we can use measurements of other diagnostic lines (e.g.\ \NII, \OIII) to quantify the fraction of H$\rm\alpha$ emission originating from H\,{\sc ii} regions as a function of galactocentric radius. Using the nebula catalogue introduced in \citet{santoro22} and described in more detail in Groves et al.~(in prep.), we find that more than 80\,\% of the H$\rm\alpha$ emission from discrete sources comes from H\,{\sc ii} regions, except for in a few galactic centres, which are not included in our analysis. This calculation does not account for H$\rm\alpha$ emission from the diffuse ionised gas (DIG), but \citet{belfiore22} demonstrate that the majority of this emission also originates from young stars, and in any case this diffuse emission is largely filtered out of our maps by the Fourier filtering described in Section~\ref{sec:method}. The PHANGS--H$\rm\alpha$ sample consists of 65 galaxies, of which 36 galaxies were observed by the du Pont 2.5-m telescope at the Las Campanas Observatory, and 32 galaxies by the Wide Field Imager (WFI) instrument at the MPG-ESO 2.2-m telescope at the La Silla Observatory, including three galaxies that overlap with the du Pont 2.5-m telescope targets. For the overlapping galaxies, we use the observations with the best angular resolution. NGC\,1097 is not included in PHANGS--H$\rm\alpha$ and therefore we use the H$\rm\alpha$ map from SINGS \citep{kennicutt03}. This observation was carried out using the CTIO 1.5m telescope with the CFCCD imager. For all the galaxies, a correction for the Milky Way dust extinction is applied following \citet{schlafly11} and an extinction curve with $R_{\rm V}=3.1$ \citep{fitzpatrick99}. We note that we do not correct for internal extinction caused by gas and dust surrounding the young stars in these H$\rm\alpha$ maps. In \citet{haydon20_ext}, we have addressed the potential impact of internal extinction on our timescale measurements, finding that it is negligible for kpc-scale molecular gas surface densities below $20\,M_{\odot}\,\rm pc^{-2}$. Most of the galaxies fall below this threshold (see Figure~\ref{fig:exsig}). However, for those with higher molecular gas surface densities, extinction may decrease the measured molecular cloud lifetimes and feedback time-scales. We remove the contamination due to the \NII lines by assuming an intensity ratio $I({\normalfont\textsc{N\,ii}})/I({\rm H}\alpha)=0.3$ \citep{kreckel16, kreckel19,santoro22}. This correction factor for \NII contamination is known to vary with galaxy mass, as well as within a galaxy with a large-scale metallicity gradient. We note that galaxy-to-galaxy variations do not affect our time-scale measurements, because we use the ratio between the flux on GMC scales and the galactic average when constraining the time-scales and this correction factor cancels out. By contrast, variations within a galaxy can potentially affect the measured deviations of flux ratios and thus the inferred time-scales. However, we expect this effect to be small as long as a random distribution of peaks in terms of radial distances can be assumed, such that any variations in \NII correction average out. The typical resolution of the H$\rm\alpha$ maps is $\sim1\arcsec$. Razza et al. (in prep.) have measured the size of the point spread function (PSF) for each image and find that it is close enough to be approximated as Gaussian. For more detailed information about the image reduction process, we refer readers to the survey paper by A. Razza et al. (in prep.) and our previous papers using a subset of the same data set \citep{schinnerer19, chevance20, pan22}.

\subsection{Homogenization of maps to common pixel grid and masking}\label{subsec:data_regrid}

In our analysis, we require gas and SFR tracer maps of a given galaxy to have the same pixel grid. Therefore, for each galaxy, we reproject both gas and SFR tracer images to share a common astrometric grid, choosing to work with the astrometry of whichever image has the coarser pixel size. When the map that is being transformed has a finer resolution than the reference map, we first convolve the map with a Gaussian kernel to the resolution of the reference map to avoid introducing artifacts. If the map being reprojected already has a coarser resolution, no additional step is performed at this stage and both maps will be convolved to similar resolutions at a later stage during the analysis. This working resolution for each galaxy is listed as $l_{\rm ap, min}$ in Table~\ref{tab:input}. This implies that we do not homogenise the resolution across the survey but work at the best available resolution. 

Due to the limited field of view of the CO maps, we restrict our analysis to regions where CO observations have been made. For most galaxies, we mask the galaxy centre because these regions are crowded and we cannot separate distinct regions at our working resolution. The radius of the galactic centre mask is listed in Table~\ref{tab:input}. Following the prescription by \citet{kim21}, we also mask some of very bright molecular gas or SFR tracer peaks identified within a galaxy. This is necessary because our method utilises small-scale variation of the gas-to-SFR flux ratios to constrain the evolutionary timeline and implicitly assumes that this timeline is well-sampled by the ensemble of gas and SFR peaks identified. Therefore, by construction, bright peaks constituting a significant fraction of the total flux might bias our results (also see \citealp{kruijssen18}). In our previous work \citep{chevance20, kim21}, these regions were found to correspond to super-luminous regions like 30\,Doradus in the LMC, or regions located at the intersection of a spiral arm and the co-rotation radius (e.g., the headlight cloud in NGC\,0628, \citealp{herrera20}). In order to find these potential overly bright regions, we first sort the peaks that are identified within our method using \textsc{Clumpfind} (see Section~\ref{sec:method}), by descending intensity. Then, we look for any gaps, which we define to exist when the $\rm n^{th}$ peak is more than twice as bright as the $\rm (n+1)^{th}$ peak. Whenever such a gap is found, we mask all the peaks that are brighter than the $\rm (n+1)^{th}$ brightest peak. This typically results in a maximum of three masked peaks in any particular galaxy.

\section{Method}\label{sec:method}

In this section we provide a brief description of our analysis method (formalised in the \textsc{Heisenberg}\footnote{https://github.com/mustang-project/Heisenberg} code) and explain the main input parameters used. A detailed explanation of the concept can be found in \citet{kruijssen14}, the presentation and validation of the code and a full description of its input parameters in \citet{kruijssen18}, and applications of the method to observed galaxies in \citet{kruijssen19}, \citet{chevance20}, \citet{chevance20_fb}, \citet{haydon20_ext}, \citet{ward20_HI}, \citet{zabel20}, and \citet{kim21}.

Our method makes use of the observational fact that galaxies are composed of numerous GMCs and star-forming regions, which are spatially decorrelated at small scale while being tightly correlated on galactic scale, defining the well-known ``star formation relation'' (e.g., \citealp{kennicutt98}). This decorrelation was first pointed out by \citet{schruba10}, and is inevitable if the CO and H$\rm\alpha$-emitting phases represent temporally-distinct stages of the GMC lifecycle \citep{kruijssen14}: the GMCs and star-forming regions represent instantaneous manifestations of individual regions undergoing independent evolution, during which molecular clouds assemble, form stars, and get disrupted by stellar feedback, only leaving young stellar regions to be detected without molecular gas. 

In order to translate this observed decorrelation between gas and young stars into their evolutionary lifecycle, we first identify emission peaks in gas and SFR tracer maps using \textsc{Clumpfind} \citep{williams94}. This algorithm finds peaks by contouring the data for a set of flux levels, which are spaced by an interval of $\rm{\delta}log_{10}\mathcal{F}$ and are spread over a flux range ($\rm{\Delta}log_{10}\mathcal{F}$) below the maximum flux level. We then reject peaks that contain less than $N_{\rm pix,min}$ pixels. Around each identified peak, we then place apertures of various sizes ranging from cloud scales ($l_{\rm ap, min}$) to galactic scales ($l_{\rm ap, max}$) and measure the relative changes of the gas-to-SFR tracer flux ratio, compared to the galactic average, at each given aperture size. We set the number of aperture sizes to be $N_{\rm ap}=15$.

We then fit an analytical function \citep[see Sect.~3.2.11 of][]{kruijssen18} to the measured flux ratios, which assumes that the measured flux ratios reflect the superposition of independently evolving regions and depend on the relative durations of the successive phases of the cloud and star formation timeline, as well as on the typical separation length between independent regions ($\lambda$). The absolute durations of the successive phases are obtained by scaling the resulting constraints on time-scale ratios by a reference time-scale ($t_{\rm ref}$). Here, we use the duration of the isolated H$\alpha$ emitting phase as $t_{\rm ref}$ ($\rm \approx 4.3\,Myr$). This value is appropriate for a delta-function star formation history and thus does not correspond to the full duration of the H$\alpha$ emitting phase if the age spread is non-zero (which is accounted for, see below). This $t_{\rm ref}$ is dependent on metallicity and listed for each galaxy in Table~\ref{tab:input}, and was calibrated by \citet{haydon20, haydon20_ext} using the stellar population synthesis model \textsc{Slug2} \citep{dasilva12,dasilva14, krumholz15}. Therefore, our fitted model is described by three independent non-degenerate parameters: the cloud lifetime ($t_{\rm CO}$), the phase during which both molecular gas and SFR tracers overlap ($t_{\rm fb}$), and the characteristic separation length between independent regions ($\lambda$). The overlapping time-scale represents the duration from when massive star-forming regions start emerging to when the surrounding molecular gas has been completely removed or dissociated by feedback, and is therefore often referred to as the feedback time-scale. The SFR tracer-emitting time-scale ($t_{\rm H\alpha}$) then simply follows as $t_{\rm fb}+t_{\rm ref}$, where the addition of $t_{\rm fb}$ allows for the presence of an age spread in individual regions.

\begin{table*}
\begin{center}
\caption{Main input parameters used in our analysis of each galaxy. Other parameters not listed here are set to the default value from \citet{kruijssen18}.} \label{tab:input}
\setlength\tabcolsep{2pt}
\begin{threeparttable}
\begin{adjustbox}{width=\textwidth}
\begin{tabular}{lccccccccccccccccc}
\hline
Galaxy&$r_{\rm min}$ & $l_{\rm ap, min}$  & $l_{\rm ap, max}$ & $N_{\rm ap}$ & $\rm N_{pix, min}$ & $\Delta\rm log_{10}\mathcal{F}_{\rm CO}$   & $\delta\rm log_{10}\mathcal{F}_{\rm CO}$ & $\Delta\rm log_{10}\mathcal{F}_{\rm H\alpha}$ & $\delta\rm log_{10}\mathcal{F}_{\rm H\alpha}$ & $t_{\rm ref}$ & $t_{\rm ref, errmin}$ &$t_{\rm ref, errmax}$ & SFR  & $\sigma$(SFR) & $\rm log_{10}~\alpha_{CO}$ & $\sigma_{\rm rel}(\rm \alpha_{CO})$ & $n_{\lambda}$ \\
& [kpc]&[pc] &[pc] & &  &   & &  & & [Myr] &[Myr]  &[Myr]  & [$M_{\rm \odot}\pyr$] & [$ M_{\rm\odot}\rm\pyr$]&  & &  \\
\hline
IC1954&0.3&132&3000&15&7&2.0&0.05&2.5&0.05&4.32&0.23&0.09&0.29&0.06&0.83&0.5&14\\
IC5273&0.8&154&3000&15&6&2.0&0.05&2.5&0.05&4.32&0.23&0.09&0.37&0.07&0.83&0.5&12\\
NGC0628&0.7&54&3000&15&20&1.4&0.10&2.8&0.10&4.28&0.22&0.08&0.76&0.15&0.76&0.5&13\\
NGC0685&0.0&170&3000&15&10&2.0&0.05&3.0&0.05&4.32&0.23&0.09&0.42&0.08&0.82&0.5&13\\
NGC1087&0.5&144&3000&15&20&1.5&0.10&3.0&0.05&4.31&0.23&0.09&1.02&0.20&0.80&0.5&13\\
NGC1097&1.9&137&3000&15&7&2.1&0.05&2.1&0.05&4.24&0.22&0.07&1.03&0.21&0.67&0.5&14\\
NGC1300&2.7&123&3000&15&8&1.5&0.05&1.8&0.05&4.22&0.21&0.06&0.67&0.13&0.64&0.5&11\\
NGC1365&5.4&174&3000&15&9&2.7&0.05&2.5&0.05&4.25&0.22&0.07&1.17&0.23&0.69&0.5&13\\
NGC1385&0.0&125&3000&15&5&1.7&0.05&3.0&0.05&4.29&0.23&0.08&1.82&0.36&0.77&0.5&14\\
NGC1433&3.4&110&3000&15&10&1.9&0.05&1.8&0.05&4.23&0.22&0.06&0.45&0.09&0.65&0.5&12\\
NGC1511&0.3&196&6000&15&15&1.8&0.05&3.0&0.05&4.33&0.23&0.09&1.52&0.30&0.84&0.5&12\\
NGC1512&4.0&110&3000&15&8&1.2&0.05&2.0&0.05&4.24&0.22&0.07&0.32&0.06&0.68&0.5&11\\
NGC1546&0.4&217&3000&15&3&2.2&0.05&3.0&0.05&4.29&0.22&0.08&0.61&0.12&0.76&0.5&15\\
NGC1559&0.0&201&6000&15&8&2.8&0.05&2.3&0.15&4.29&0.23&0.08&4.12&0.82&0.78&0.5&13\\
NGC1566&0.5&115&3000&15&5&1.8&0.05&2.6&0.05&4.25&0.22&0.07&3.02&0.60&0.69&0.5&12\\
NGC1672&2.7&212&3000&15&10&1.9&0.03&3.0&0.10&4.26&0.22&0.07&2.09&0.42&0.72&0.5&14\\
NGC1792&1.5&233&3000&15&7&1.8&0.05&2.5&0.05&4.27&0.22&0.08&2.90&0.58&0.73&0.5&14\\
NGC1809&0.0&186&6000&15&6&2.0&0.05&3.0&0.05&4.33&0.23&0.09&0.18&0.04&0.84&0.5&13\\
NGC2090&0.7&112&3000&15&10&2.0&0.05&2.5&0.05&4.32&0.23&0.09&0.15&0.03&0.83&0.5&12\\
NGC2283&0.0&102&3000&15&17&1.4&0.05&3.0&0.05&4.35&0.22&0.09&0.49&0.10&0.88&0.5&13\\
NGC2835&0.0&74&3000&15&15&1.2&0.05&3.0&0.05&4.32&0.23&0.09&0.39&0.08&0.83&0.5&12\\
NGC2997&2.0&132&3000&15&30&1.8&0.05&2.8&0.05&4.25&0.22&0.07&2.61&0.52&0.68&0.5&12\\
NGC3059&0.8&150&5000&15&20&2.0&0.05&3.5&0.05&4.28&0.22&0.08&1.33&0.27&0.75&0.5&10\\
NGC3351&2.0&84&3000&15&12&1.8&0.05&2.8&0.05&4.25&0.22&0.07&0.20&0.04&0.69&0.5&12\\
NGC3507&2.0&176&7000&15&7&1.8&0.05&3.0&0.10&4.25&0.22&0.07&0.60&0.12&0.70&0.5&12\\
NGC3511&0.8&240&6000&15&6&2.0&0.05&2.5&0.05&4.33&0.23&0.09&0.66&0.13&0.84&0.5&13\\
NGC3596&0.3&75&3000&15&10&2.0&0.05&3.5&0.10&4.38&0.22&0.08&0.23&0.05&0.93&0.5&14\\
NGC3627&1.1&121&3000&15&12&3.2&0.10&3.1&0.10&4.25&0.22&0.07&2.71&0.54&0.70&0.5&13\\
NGC4254&0.2&125&3000&15&12&2.0&0.05&3.5&0.05&4.27&0.22&0.08&2.62&0.52&0.74&0.5&13\\
NGC4298&0.6&160&3000&15&7&2.5&0.05&3.5&0.05&4.29&0.23&0.08&0.38&0.08&0.77&0.5&14\\
NGC4303&1.2&156&3000&15&5&1.6&0.10&2.5&0.05&4.24&0.22&0.07&3.61&0.72&0.68&0.5&13\\
NGC4321&1.0&139&3000&15&10&1.6&0.15&1.6&0.25&4.24&0.22&0.07&2.20&0.44&0.68&0.5&12\\
NGC4496A&0.0&118&3000&15&10&2.5&0.05&3.0&0.05&4.36&0.22&0.09&0.18&0.04&0.90&0.5&12\\
NGC4535&3.0&141&7200&15&10&2.5&0.05&2.7&0.10&4.24&0.22&0.07&0.76&0.15&0.67&0.5&12\\
NGC4540&0.0&112&3000&15&3&3.0&0.05&3.0&0.05&4.33&0.23&0.09&0.14&0.03&0.85&0.5&14\\
NGC4548&1.4&150&3000&15&8&1.8&0.05&2.5&0.05&4.24&0.22&0.07&0.22&0.04&0.67&0.5&14\\
NGC4569&1.6&220&5000&15&3&1.4&0.03&2.0&0.03&4.23&0.21&0.06&0.68&0.14&0.65&0.5&13\\
NGC4571&0.4&128&3000&15&7&2.0&0.05&3.0&0.40&4.30&0.23&0.08&0.17&0.03&0.78&0.5&12\\
NGC4654&2.1&243&5000&15&15&2.0&0.05&3.2&0.05&4.28&0.06&0.32&2.31&0.46&0.74&0.5&12\\
NGC4689&1.7&111&3000&15&15&1.4&0.05&3.0&0.05&4.27&0.22&0.08&0.41&0.08&0.74&0.5&13\\
NGC4731&0.0&149&3000&15&5&2.5&0.16&2.5&0.05&4.36&0.22&0.09&0.19&0.04&0.90&0.5&13\\
NGC4781&0.0&100&3000&15&15&1.8&0.05&1.6&0.05&4.32&0.23&0.09&0.40&0.08&0.83&0.5&12\\
NGC4941&1.4&149&3000&15&10&1.6&0.05&2.5&0.05&4.28&0.22&0.08&0.13&0.03&0.75&0.5&12\\
NGC4951&0.5&167&5000&15&10&2.0&0.05&2.4&0.05&4.34&0.23&0.09&0.17&0.03&0.86&0.5&12\\
NGC5042&0.7&134&3000&15&10&2.5&0.10&2.5&0.05&4.36&0.22&0.09&0.22&0.04&0.89&0.5&12\\
NGC5068&0.0&32&3000&15&40&1.6&0.30&3.4&0.10&4.43&0.21&0.08&0.18&0.04&1.03&0.5&10\\
NGC5134&1.2&124&3000&15&10&1.3&0.05&3.4&0.05&4.28&0.22&0.08&0.30&0.06&0.75&0.5&12\\
NGC5248&1.2&113&3000&15&10&2.0&0.10&2.4&0.03&4.27&0.22&0.08&1.35&0.27&0.74&0.5&13\\
NGC5530&0.0&105&3000&15&15&1.6&0.05&2.0&0.05&4.30&0.23&0.08&0.33&0.07&0.79&0.5&10\\
NGC5643&1.1&86&3000&15&15&1.4&0.10&2.5&0.10&4.26&0.22&0.07&1.11&0.22&0.72&0.5&11\\
NGC6300&1.7&85&3000&15&15&1.2&0.15&2.2&0.05&4.25&0.22&0.07&0.63&0.13&0.69&0.5&13\\
NGC6744&3.0&78&3000&15&10&1.4&0.05&2.6&0.30&4.28&0.22&0.08&0.49&0.10&0.75&0.5&12\\
NGC7456&0.0&206&5000&15&3&4.0&0.05&3.5&0.05&4.48&0.20&0.07&0.10&0.02&1.12&0.5&11\\
NGC7496&1.3&169&5000&15&5&3.0&0.05&3.2&0.05&4.27&0.22&0.08&0.53&0.11&0.73&0.5&13\\
\hline
\end{tabular}
\end{adjustbox}
\end{threeparttable}
\end{center}
\end{table*}

To focus our analysis specifically on the peaks of emission, we follow \citet{kruijssen19}, \citet{chevance20} and \citet{kim21} and apply a Fourier filter to remove the large-scale structure from both maps using the method presented in \citet{hygate19}. In the H$\rm\alpha$ maps, this will suppress diffuse ionised gas emission, which often forms a large halo around H\,{\sc ii} regions \citep[see e.g.][]{haffner09, rahman11, belfiore22} and add a large-scale reservoir of emission that does not originate from peaks within the aperture. In the CO maps, the operation isolates clumps of emission, likely to be the individual massive clouds or complexes, most directly associated with the massive H\,{\sc ii} regions that we identify. In practice, we filter out emission on spatial scales larger than $n_{\lambda}$ times the typical separation length between independent star-forming regions ($\lambda$; constrained in our method), using a Gaussian high-pass filter in Fourier space. We choose the smallest possible value of $n_{\lambda}$ (see Table~\ref{tab:input}), while ensuring that flux loss from the compact regions is less than 10\%, following the prescription by \citet{hygate19}, \citet{kruijssen19}, \citet{chevance20}, and \citet{kim21}. Because the width of the filter depends on the region separation length, which is a quantity constrained by the analysis, this procedure must be carried out iteratively until a convergence condition is reached. This condition is defined as a change of the measured $\lambda$ by less than 5\% for three consecutive iterations.

Using this method, we also derive other physical properties, such as the compact molecular gas surface density ($\Sigma_{\rm H_{2}}^{\rm compact}$), SFR surface density of the analysed region ($\Sigma_{\rm SFR}$), depletion time of the compact molecular gas ($t_{\rm dep}^{\rm compact}$, which assumes that all of the SFR results from the compact molecular gas component)\footnote{We also derive molecular gas surface density (including the diffuse component; $\Sigma_{\rm H_{2}}$) and depletion time for all the molecular gas $t_{\rm dep}$}, and integrated star formation efficiency achieved by the average star-forming region ($\epsilon_{\rm sf}$). To derive these properties, we adopt a \cotwo-to-$\rm H_{2}$ conversion factor ($\rm \alpha_{CO}$) and use the global SFR, which includes a contribution from diffused ionised gas and a correction for the internal extinction. We adopt a metallicity-dependent $\rm \alpha_{CO}$ following \citet{sun20}, expressed as 
\begin{equation}
    \alpha_{\rm CO} = 4.35\,Z'^{-1.6}/R_{21}~\Msun~({\rm K}~\kms~\pc^2)^{-1}~,
\end{equation}\label{eq:conv_gas} 
where $ R_{\rm21}$ is the \cotwo-to-\coone line ratio (0.65; \citealp{leroy13, denbrok21, leroy21_lineratio}) and $Z'$ is the CO luminosity-weighted metallicity in units of the solar value. To calculate this, we adopt the metallicity from the global stellar mass-metallicity relation \citep{sanchez19} at the effective radius ($\rm R_{eff}$) and assume a fixed radial metallicity gradient within the galaxy ($\rm-0.1dex/R_{\rm eff}$; \citealp{sanchez14}). The global SFR, accounting for extinction, is adopted from \citet{leroy19} and is measured by combining maps from the Galaxy Evolution Explorer (GALEX) far ultraviolet band (155\,nm) and the Wide-field Infrared Survey Explorer (WISE) W4 band (22\,$\mu$m), convolved to 15\arcsec\ angular resolution. Here, we only consider star formation within the analysed region, excluding galactic centres and bright regions that are masked in our analysis. We assume a conservative uncertainty for $\rm \alpha_{CO}$ of 50\% and a typical SFR uncertainty of 20\%, denoted as $\rm\sigma_{rel}(\alpha_{CO})$ and $\rm\sigma(SFR)$, respectively in Table~\ref{tab:input}.

The main input parameters explained above are listed in Table~\ref{tab:input} for each galaxy. We use the distance, inclination angle, and position angle listed in Table~\ref{tab:sample}. For other parameters not mentioned here, we use the default values as listed in \citet{kruijssen18}, which are related to the fitting process and error propagation.

\section{Evolutionary timeline from molecular gas to exposed young stellar regions}\label{sec:results}

In this section, we present our results obtained from the application of our statistical method to CO and H$\rm\alpha$ observations as tracers of molecular gas and SFR for the 54 PHANGS galaxies described in Section~\ref{sec:data}. We note that our results include the re-analysis of 8 galaxies from \citet[][excluding M51, which is not in our galaxy sample]{chevance20}. Here, we use an updated H$\rm\alpha$ map for NGC3627, which is reduced with a better continuum subtraction (A. Razza in prep.), as well as improved CO data products for all of the re-analysed galaxies. In addition, we introduced an additional step where, for each CO-H$\rm\alpha$ pair of maps, we convolve one map to match the resolution of the other map, \textit{before} reprojecting it to the coarsest pixel grid (as explained in Section~\ref{subsec:data_regrid}). Despite these changes, our results agree to within the 1$\sigma$ uncertainties with those from \citet{chevance20}.

\begin{figure*}
\includegraphics[scale=0.34]{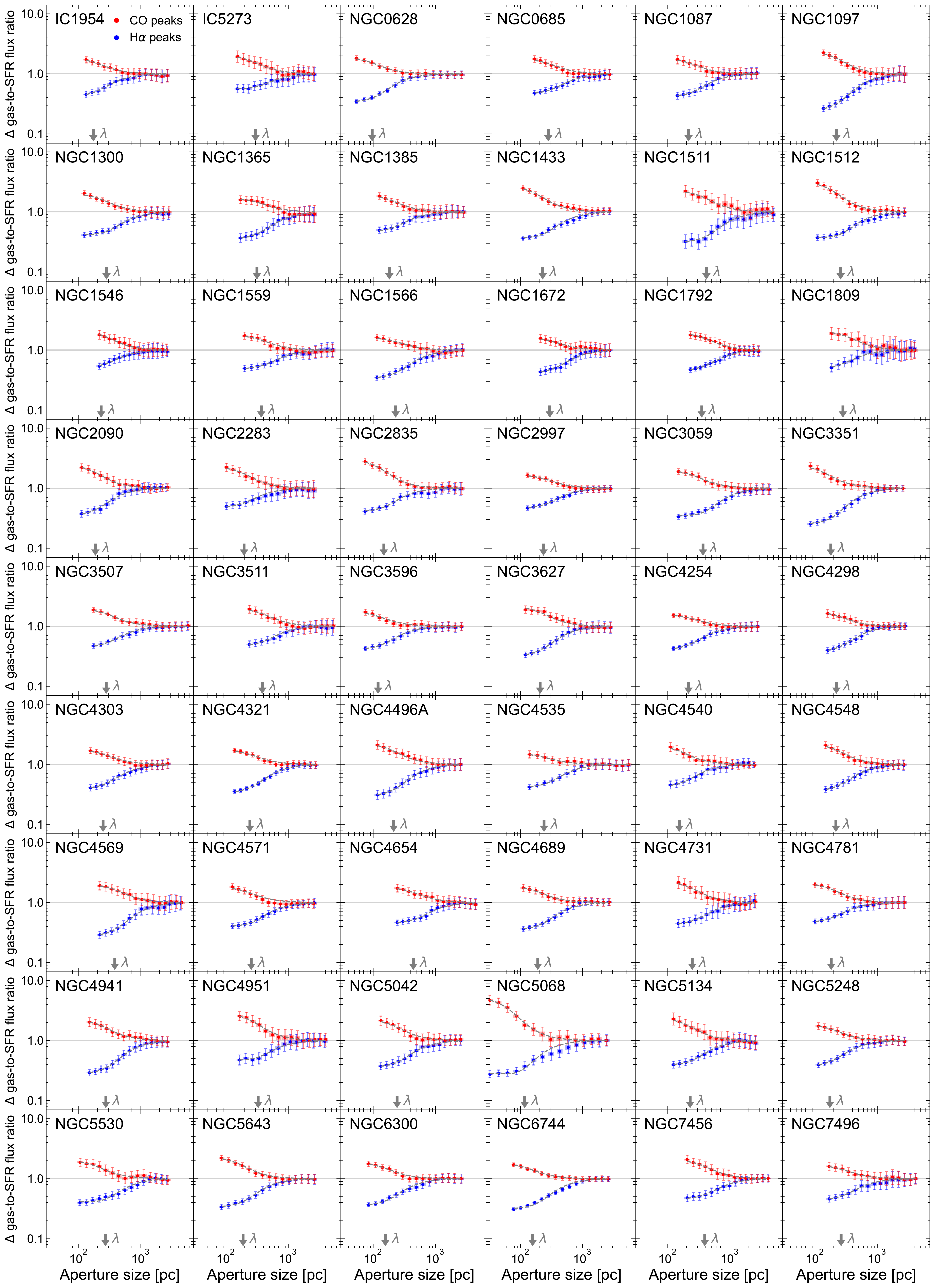}
\caption{Measured variation of the gas-to-SFR tracer flux ratio (CO-to-H$\rm\alpha$) relative to the galactic average as a function of the size of apertures placed on CO (red) and H$\rm\alpha$ (blue) emission peaks. The error bars indicate a 1$\sigma$ uncertainty on each individual measurement whereas the  shaded area around each error bar represents the effective 1$\sigma$ uncertainty used in the fitting process, which takes into account the covariance between measurements. The galactic average is shown as the solid horizontal line and the dashed line indicates our best-fitting model. The best-fitting region separation length ($\lambda$) is indicated in each panel with an arrow and other constrained parameters are listed in Table~\ref{tab:result}. } \label{fig:tuningforks}
\end{figure*}

\begin{figure*}
\includegraphics[scale=0.77]{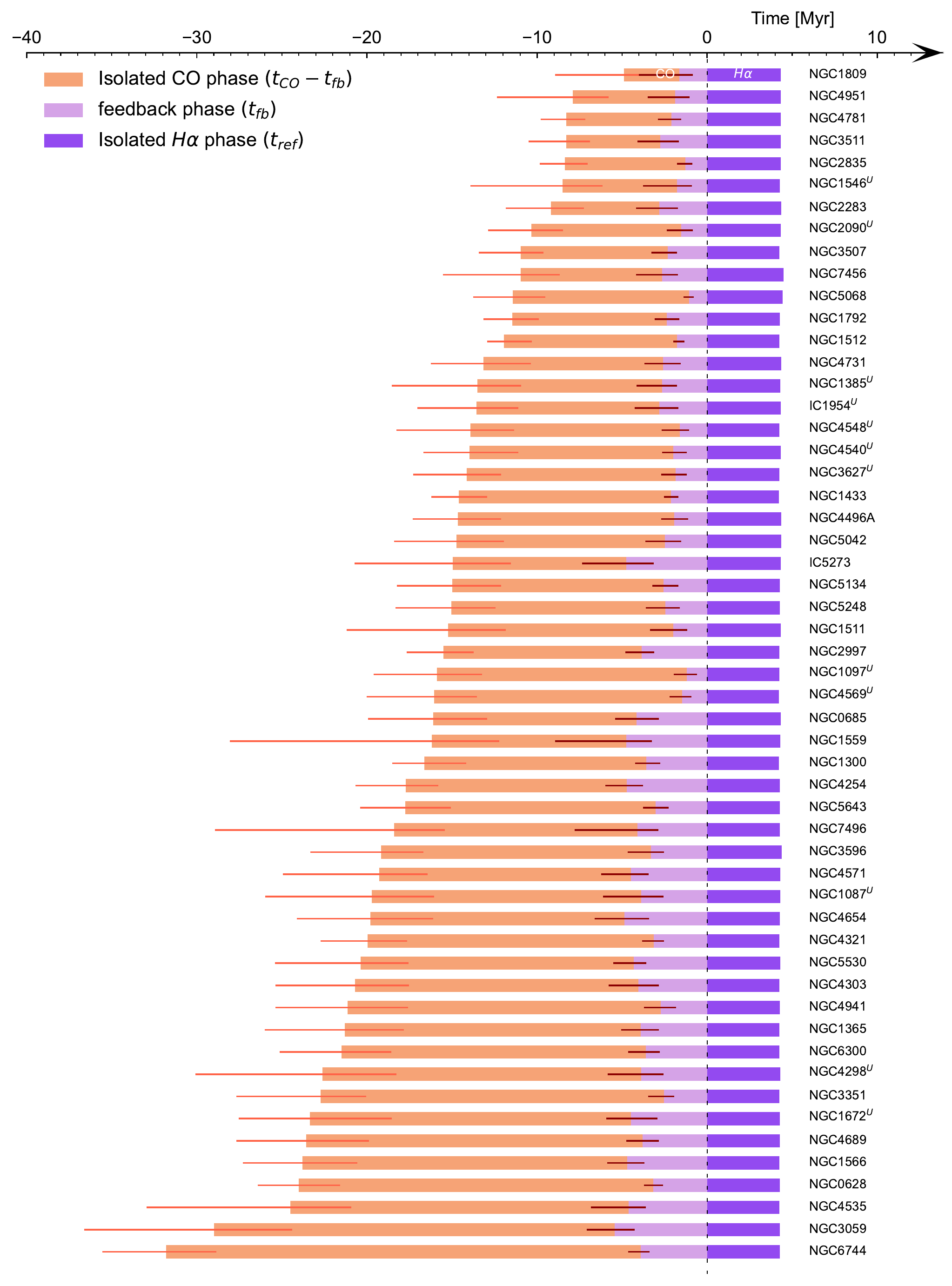}
\caption{Evolutionary timeline of GMCs from molecular gas assembly to feedback-driven dispersal, and H\,{\sc ii} regions free of molecular gas, ordered by increasing cloud lifetime from top to bottom. Going from left to right, GMCs are initially only visible in CO (orange, for a duration of $t_{\rm CO}-t_{\rm fb}$). Towards the end of this phase, massive star formation has taken place, generating spatially coincident  H$\rm\alpha$ emission (light purple, for a duration of $t_{\rm fb}$). Finally, violent feedback from young stars has completely cleared the surrounding molecular gas, only leaving H$\rm\alpha$ emission to be detected, without any associated CO emission (dark purple, with a duration of $t_{\rm ref}\approx4.3$\,Myr; see Section~\ref{sec:method}). The error bars on the left and in the middle indicate the uncertainties on $t_{\rm CO}$ and $t_{\rm fb}$, respectively. $U$ indicates galaxies with only upper limit constraints on $t_{\rm fb}$.} \label{fig:timeline}
\end{figure*}

\begin{figure*}
\includegraphics[scale=0.48]{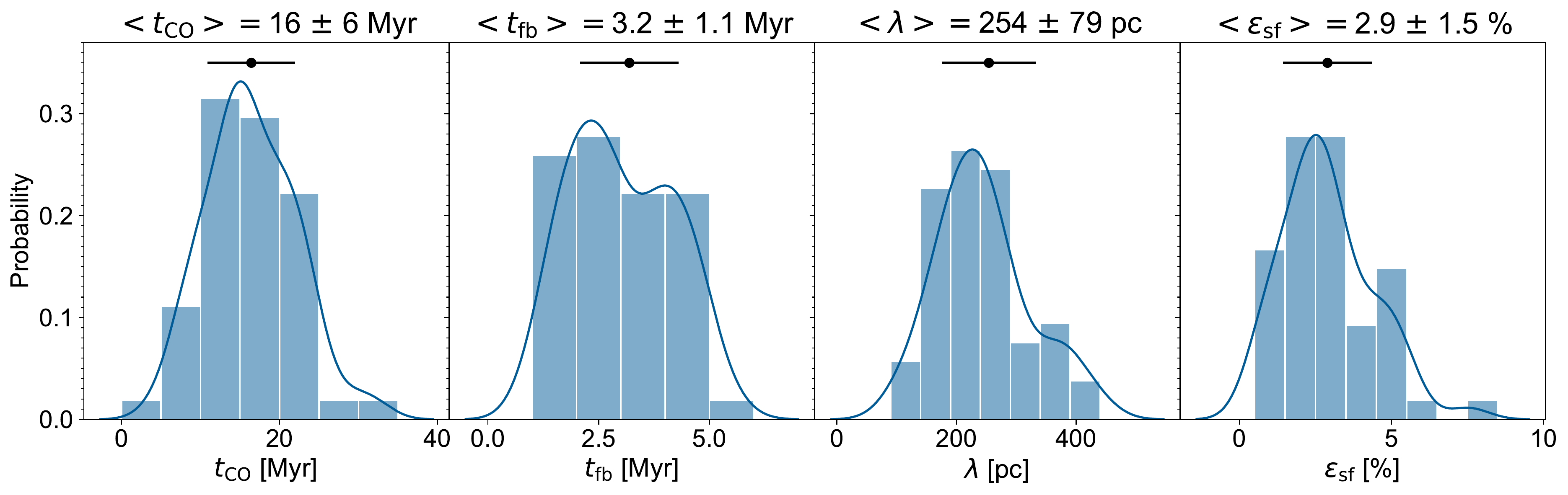}
\caption{Distributions of the main physical quantities constrained across 54 PHANGS galaxies. From left to right, we show a Gaussian kernel density estimate of the cloud lifetime ($t_{\rm CO}$), the feedback time-scale ($t_{\rm fb}$), the average separation length between independent star-forming regions ($\lambda$), and the integrated star formation efficiency ($\epsilon_{\rm sf}$). For each histogram, the mean (black dot) and 16-84\% range (error bar) of the data, are indicated at the top of the panel (excluding $t_{\rm fb}$ and $\lambda$ measurements where only upper limits are constrained; see Table~\ref{tab:result}).}  \label{fig:histo}
\end{figure*}

Figure~\ref{fig:tuningforks} shows the variations of the gas-to-SFR tracer flux ratios measured within apertures centred on CO and H$\rm\alpha$ peaks relative to the galactic average, as a function of the aperture size, together with our best-fitting model for each galaxy. For all the galaxies in our sample, the measured flux ratios increasingly diverge from the galactic average as the size of the aperture decreases (from $\sim$1\,kpc to $\sim$50\,pc), both for apertures focused on gas and SFR tracer peaks. This demonstrates a universal spatial decorrelation between molecular gas and young stellar regions on the cloud scale, at the sensitivity limits of our data. In Table~\ref{tab:result}, we present the best-fitting free parameters constrained by the model fit to each galaxy, as well as additional quantities that can be derived from our measurements: the feedback outflow velocity (\vfb; see Section~\ref{sssec:vfb}), the integrated cloud-scale star formation efficiency (\esf; see Section~\ref{sssec:esf}), and the fractions of diffuse molecular and ionised gas ($f_{\rm diffuse}^ {\rm CO}$ and $f_{\rm diffuse}^ {\rm  H\alpha}$; see Section~\ref{sssec:diffuse}), which are determined during our diffuse emission filtering process described in Section~\ref{sec:method}. Figure~\ref{fig:timeline} shows an illustration of the resulting evolutionary lifecycles of GMCs in our galaxy sample, from the assembly of molecular gas to the feedback phase powered by massive star formation, and finally to exposed young stellar regions. Star formation regions emit in only CO in the beginning, then also in H$\rm\alpha$ after the massive star-forming region has become partially exposed, and finally only in H$\rm\alpha$ after cloud disruption. In Figure~\ref{fig:histo}, we show the distributions of our main measurements across the galaxy sample.  

\begin{table*}
\begin{center}
\caption{Physical quantities constrained by applying the method presented in Section~\ref{sec:method}, describing the evolution of molecular clouds to exposed young stellar regions. The columns list the cloud lifetime ($t_{\rm CO}$), the feedback time-scale ($t_{\rm fb}$), the H$\rm\alpha$-emitting time-scale ($t_{\rm H\alpha}=t_{\rm fb}+t_{\rm ref}$), the characteristic region separation length ($\lambda$), the feedback velocity (\vfb), the integrated star formation efficiency (\esf), the diffuse emission fraction in the molecular gas and SFR tracer maps ($f_{\rm diffuse}^ {\rm CO}$ and $f_{\rm diffuse}^ {\rm  H\alpha}$, respectively), and the depletion time-scale of the compact molecular gas ($t_{\rm dep}^{\rm compact}$), as well as the depletion time-scale for total molecular gas ($t_{\rm dep}$), which includes the diffuse component. For some galaxies, only a 1$\sigma$ upper limit can be obtained on $t_{\rm fb}$, $t_{\rm H\alpha}$, and $\lambda$, because the independent star-forming regions are not sufficiently resolved (see Appendix~\ref{sec:accuracy}). } \label{tab:result}
\begin{threeparttable}
\setlength\tabcolsep{8pt}
\begin{tabular}{lcccccccccc}
\hline
Galaxy & $t_{\rm CO}$  & $t_{\rm fb}$ & $t_{\rm H\alpha}$ & $\lambda$ &\vfb\ &\esf\ &$f_{\rm diffuse}^ {\rm CO}$&$f_{\rm diffuse}^ {\rm  H\alpha}$&$t_{\rm dep}^{\rm compact}$ &$t_{\rm dep}$ \\ 
&[Myr] &[Myr]&[Myr] &[pc] &  [km s$^{-1}$]& [per~cent]&&&[Gyr]&[Gyr]\\
\hline
IC1954
& $13.6_{-2.5}^{+3.5}$& $<\,4.3$& $<\,8.6$& $<\,225$& $18.6_{-5.2}^{+10.0}$& $1.9_{-0.8}^{+1.6}$& $0.37_{-0.05}^{+0.06}$& $0.72_{-0.02}^{+0.02}$& $0.7_{-0.3}^{+0.5}$& $2.2_{-0.9}^{+1.6}$\\
IC5273
& $14.9_{-3.4}^{+5.8}$& $4.7_{-1.6}^{+2.6}$& $9.1_{-1.6}^{+2.6}$& $299_{-68}^{+166}$& $14.2_{-4.2}^{+5.8}$& $4.9_{-2.1}^{+4.9}$& $0.47_{-0.03}^{+0.04}$& $0.65_{-0.03}^{+0.05}$& $0.3_{-0.1}^{+0.2}$& $0.9_{-0.4}^{+0.6}$\\
NGC0628
& $24.0_{-2.4}^{+2.4}$& $3.2_{-0.6}^{+0.5}$& $7.4_{-0.6}^{+0.6}$& $96_{-10}^{+13}$& $8.0_{-1.0}^{+1.4}$& $5.5_{-2.3}^{+4.0}$& $0.52_{-0.03}^{+0.03}$& $0.38_{-0.03}^{+0.03}$& $0.4_{-0.2}^{+0.3}$& $1.6_{-0.7}^{+1.2}$\\
NGC0685
& $16.1_{-3.2}^{+3.8}$& $4.2_{-1.3}^{+1.3}$& $8.5_{-1.4}^{+1.3}$& $280_{-45}^{+44}$& $18.1_{-4.0}^{+7.1}$& $3.6_{-1.6}^{+2.9}$& $0.13_{-0.05}^{+0.05}$& $0.59_{-0.03}^{+0.02}$& $0.4_{-0.2}^{+0.3}$& $1.0_{-0.4}^{+0.7}$\\
NGC1087
& $19.7_{-3.7}^{+6.3}$& $<\,6.1$& $<\,10.5$& $<\,290$& $15.7_{-4.9}^{+6.4}$& $5.3_{-2.3}^{+4.6}$& $0.45_{-0.03}^{+0.03}$& $0.53_{-0.04}^{+0.04}$& $0.4_{-0.2}^{+0.3}$& $1.2_{-0.5}^{+0.8}$\\
NGC1097
& $15.9_{-2.7}^{+3.7}$& $<\,1.9$& $<\,6.2$& $220_{-25}^{+39}$& $52.7_{-18.8}^{+47.4}$& $2.4_{-1.0}^{+2.0}$& $0.35_{-0.03}^{+0.04}$& $0.70_{-0.02}^{+0.02}$& $0.7_{-0.3}^{+0.5}$& $1.8_{-0.8}^{+1.3}$\\
NGC1300
& $16.6_{-2.5}^{+1.9}$& $3.6_{-0.8}^{+0.7}$& $7.8_{-0.9}^{+0.7}$& $280_{-35}^{+46}$& $16.9_{-2.2}^{+4.6}$& $2.9_{-1.3}^{+2.1}$& $0.43_{-0.02}^{+0.02}$& $0.43_{-0.03}^{+0.02}$& $0.6_{-0.2}^{+0.4}$& $1.5_{-0.6}^{+1.0}$\\
NGC1365
& $21.3_{-3.5}^{+4.7}$& $3.9_{-1.1}^{+1.1}$& $8.1_{-1.1}^{+1.2}$& $315_{-65}^{+126}$& $22.0_{-3.9}^{+5.0}$& $2.4_{-1.0}^{+1.9}$& $0.34_{-0.05}^{+0.05}$& $0.51_{-0.05}^{+0.06}$& $0.9_{-0.4}^{+0.6}$& $2.5_{-1.0}^{+1.8}$\\
NGC1385
& $13.5_{-2.6}^{+5.0}$& $<\,4.1$& $<\,8.5$& $<\,238$& $20.0_{-6.3}^{+7.5}$& $4.7_{-2.0}^{+4.3}$& $0.50_{-0.03}^{+0.04}$& $0.62_{-0.03}^{+0.04}$& $0.3_{-0.1}^{+0.2}$& $0.9_{-0.4}^{+0.7}$\\
NGC1433
& $14.6_{-1.7}^{+1.6}$& $2.1_{-0.4}^{+0.4}$& $6.3_{-0.5}^{+0.5}$& $227_{-21}^{+27}$& $22.6_{-3.6}^{+5.4}$& $3.6_{-1.5}^{+2.7}$& $0.48_{-0.01}^{+0.02}$& $0.36_{-0.04}^{+0.05}$& $0.4_{-0.2}^{+0.3}$& $1.2_{-0.5}^{+0.8}$\\
NGC1511
& $15.2_{-3.4}^{+5.9}$& $2.0_{-0.8}^{+1.4}$& $6.3_{-0.8}^{+1.4}$& $417_{-96}^{+182}$& $49.7_{-18.8}^{+28.6}$& $4.4_{-1.9}^{+4.1}$& $0.55_{-0.05}^{+0.07}$& $0.83_{-0.02}^{+0.04}$& $0.3_{-0.1}^{+0.2}$& $1.5_{-0.6}^{+1.0}$\\
NGC1512
& $11.9_{-1.6}^{+1.0}$& $1.8_{-0.4}^{+0.2}$& $6.0_{-0.5}^{+0.3}$& $258_{-30}^{+37}$& $27.9_{-2.5}^{+8.3}$& $2.2_{-1.0}^{+1.5}$& $0.53_{-0.02}^{+0.02}$& $0.33_{-0.02}^{+0.02}$& $0.5_{-0.2}^{+0.4}$& $1.7_{-0.7}^{+1.2}$\\
NGC1546
& $8.5_{-2.4}^{+5.4}$& $<\,3.8$& $<\,8.1$& $<\,348$& $44.2_{-19.2}^{+32.0}$& $0.7_{-0.3}^{+0.8}$& $0.69_{-0.04}^{+0.10}$& $0.31_{-0.37}^{+1.07}$& $1.2_{-0.5}^{+0.9}$& $7.6_{-3.1}^{+5.4}$\\
NGC1559
& $16.2_{-4.0}^{+11.9}$& $4.7_{-1.5}^{+4.2}$& $9.0_{-1.5}^{+4.2}$& $369_{-63}^{+140}$& $21.1_{-9.1}^{+7.9}$& $3.3_{-1.4}^{+4.2}$& $0.43_{-0.03}^{+0.04}$& $0.28_{-0.14}^{+0.23}$& $0.5_{-0.2}^{+0.3}$& $1.4_{-0.6}^{+1.0}$\\
NGC1566
& $23.8_{-3.2}^{+3.5}$& $4.7_{-1.0}^{+1.2}$& $9.0_{-1.1}^{+1.2}$& $229_{-35}^{+50}$& $12.6_{-2.2}^{+2.8}$& $4.1_{-1.8}^{+3.1}$& $0.36_{-0.02}^{+0.02}$& $0.46_{-0.03}^{+0.03}$& $0.6_{-0.2}^{+0.4}$& $1.3_{-0.6}^{+1.0}$\\
NGC1672
& $23.4_{-4.8}^{+4.2}$& $<\,5.9$& $<\,10.2$& $<\,442$& $20.1_{-3.4}^{+7.3}$& $4.7_{-2.1}^{+3.4}$& $0.28_{-0.07}^{+0.07}$& $0.43_{-0.05}^{+0.05}$& $0.5_{-0.2}^{+0.4}$& $1.3_{-0.5}^{+0.9}$\\
NGC1792
& $11.4_{-1.5}^{+1.7}$& $2.4_{-0.7}^{+0.7}$& $6.6_{-0.8}^{+0.8}$& $352_{-58}^{+96}$& $43.8_{-7.2}^{+13.8}$& $1.3_{-0.5}^{+0.9}$& $0.49_{-0.03}^{+0.04}$& $0.67_{-0.03}^{+0.03}$& $0.9_{-0.4}^{+0.6}$& $3.2_{-1.3}^{+2.3}$\\
NGC1809
& $4.9_{-1.0}^{+4.0}$& $1.6_{-0.8}^{+2.4}$& $6.0_{-0.8}^{+2.5}$& $280_{-83}^{+602}$& $50.8_{-23.6}^{+30.2}$& $0.8_{-0.4}^{+1.2}$& $0.34_{-0.09}^{+0.25}$& $0.62_{-0.06}^{+0.21}$& $0.6_{-0.2}^{+0.4}$& $1.6_{-0.6}^{+1.1}$\\
NGC2090
& $10.3_{-1.9}^{+2.5}$& $<\,2.4$& $<\,6.7$& $185_{-28}^{+43}$& $31.3_{-10.1}^{+23.3}$& $0.8_{-0.4}^{+0.7}$& $0.46_{-0.03}^{+0.04}$& $0.81_{-0.01}^{+0.01}$& $1.2_{-0.5}^{+0.9}$& $4.5_{-1.9}^{+3.2}$\\
NGC2283
& $9.2_{-1.9}^{+2.7}$& $2.8_{-1.1}^{+1.4}$& $7.2_{-1.1}^{+1.4}$& $195_{-40}^{+93}$& $17.3_{-4.8}^{+8.9}$& $2.6_{-1.1}^{+2.2}$& $0.23_{-0.07}^{+0.10}$& $0.56_{-0.03}^{+0.04}$& $0.4_{-0.1}^{+0.3}$& $0.9_{-0.4}^{+0.6}$\\
NGC2835
& $8.4_{-1.3}^{+1.5}$& $1.3_{-0.4}^{+0.5}$& $5.6_{-0.5}^{+0.6}$& $148_{-19}^{+31}$& $26.5_{-6.9}^{+11.7}$& $2.1_{-0.9}^{+1.6}$& $0.34_{-0.03}^{+0.04}$& $0.18_{-0.08}^{+0.10}$& $0.4_{-0.2}^{+0.3}$& $1.0_{-0.4}^{+0.7}$\\
NGC2997
& $15.5_{-1.8}^{+2.2}$& $3.8_{-0.7}^{+0.9}$& $8.1_{-0.8}^{+1.0}$& $234_{-30}^{+46}$& $16.4_{-2.6}^{+2.9}$& $3.0_{-1.3}^{+2.2}$& $0.35_{-0.03}^{+0.04}$& $0.33_{-0.04}^{+0.04}$& $0.5_{-0.2}^{+0.4}$& $1.5_{-0.6}^{+1.1}$\\
NGC3059
& $29.0_{-4.6}^{+7.7}$& $5.4_{-1.2}^{+1.6}$& $9.7_{-1.2}^{+1.7}$& $369_{-40}^{+50}$& $13.7_{-3.0}^{+3.5}$& $7.5_{-3.2}^{+6.3}$& $0.36_{-0.03}^{+0.04}$& $0.52_{-0.08}^{+0.12}$& $0.4_{-0.2}^{+0.3}$& $1.2_{-0.5}^{+0.8}$\\
NGC3351
& $22.7_{-2.7}^{+5.0}$& $2.5_{-0.6}^{+0.9}$& $6.8_{-0.7}^{+1.0}$& $179_{-20}^{+24}$& $14.4_{-3.7}^{+4.2}$& $3.4_{-1.4}^{+2.8}$& $0.49_{-0.01}^{+0.01}$& $0.23_{-0.02}^{+0.02}$& $0.7_{-0.3}^{+0.5}$& $2.0_{-0.8}^{+1.5}$\\
NGC3507
& $11.0_{-1.3}^{+2.5}$& $2.3_{-0.6}^{+1.0}$& $6.6_{-0.6}^{+1.0}$& $277_{-38}^{+91}$& $32.9_{-8.0}^{+8.1}$& $2.4_{-1.0}^{+1.9}$& $0.46_{-0.03}^{+0.03}$& $0.64_{-0.04}^{+0.05}$& $0.5_{-0.2}^{+0.3}$& $1.4_{-0.6}^{+1.0}$\\
NGC3511
& $8.3_{-1.4}^{+2.2}$& $2.8_{-1.1}^{+1.3}$& $7.1_{-1.1}^{+1.4}$& $383_{-91}^{+146}$& $37.6_{-10.3}^{+17.8}$& $0.6_{-0.2}^{+0.5}$& $0.53_{-0.03}^{+0.03}$& $0.61_{-0.03}^{+0.03}$& $1.4_{-0.6}^{+1.0}$& $4.8_{-2.0}^{+3.4}$\\
NGC3596
& $19.2_{-2.5}^{+4.2}$& $3.3_{-0.8}^{+1.4}$& $7.7_{-0.8}^{+1.4}$& $119_{-18}^{+34}$& $9.8_{-2.5}^{+2.2}$& $2.7_{-1.1}^{+2.2}$& $0.49_{-0.03}^{+0.04}$& $0.15_{-0.06}^{+0.07}$& $0.7_{-0.3}^{+0.5}$& $2.3_{-0.9}^{+1.6}$\\
NGC3627
& $14.1_{-2.0}^{+3.1}$& $<\,2.7$& $<\,7.0$& $207_{-31}^{+47}$& $31.6_{-8.4}^{+14.0}$& $1.8_{-0.8}^{+1.5}$& $0.39_{-0.04}^{+0.05}$& $0.60_{-0.03}^{+0.03}$& $0.8_{-0.3}^{+0.6}$& $2.4_{-1.0}^{+1.7}$\\
NGC4254
& $17.7_{-1.9}^{+3.0}$& $4.7_{-1.0}^{+1.3}$& $9.0_{-1.0}^{+1.3}$& $215_{-34}^{+48}$& $12.3_{-1.9}^{+2.2}$& $2.8_{-1.2}^{+2.2}$& $0.34_{-0.05}^{+0.05}$& $0.39_{-0.03}^{+0.04}$& $0.6_{-0.3}^{+0.5}$& $2.0_{-0.8}^{+1.4}$\\
NGC4298
& $22.6_{-4.4}^{+7.5}$& $<\,5.8$& $<\,10.2$& $<\,263$& $16.7_{-5.2}^{+7.4}$& $1.8_{-0.8}^{+1.6}$& $0.51_{-0.03}^{+0.03}$& $0.48_{-0.03}^{+0.03}$& $1.3_{-0.5}^{+0.9}$& $4.9_{-2.1}^{+3.5}$\\
NGC4303
& $20.7_{-3.2}^{+4.7}$& $4.0_{-1.2}^{+1.7}$& $8.3_{-1.2}^{+1.8}$& $247_{-40}^{+77}$& $17.1_{-4.1}^{+5.5}$& $5.5_{-2.3}^{+4.5}$& $0.48_{-0.02}^{+0.02}$& $0.49_{-0.03}^{+0.03}$& $0.4_{-0.2}^{+0.3}$& $1.2_{-0.5}^{+0.8}$\\
NGC4321
& $20.0_{-2.3}^{+2.8}$& $3.1_{-0.6}^{+0.7}$& $7.4_{-0.7}^{+0.7}$& $242_{-21}^{+23}$& $20.5_{-3.4}^{+4.3}$& $3.0_{-1.3}^{+2.3}$& $0.47_{-0.01}^{+0.01}$& $0.45_{-0.02}^{+0.02}$& $0.7_{-0.3}^{+0.5}$& $2.1_{-0.9}^{+1.5}$\\
NGC4496A
& $14.6_{-2.5}^{+2.7}$& $1.9_{-0.8}^{+0.8}$& $6.3_{-0.9}^{+0.8}$& $213_{-37}^{+52}$& $27.2_{-6.9}^{+17.9}$& $2.4_{-1.0}^{+1.8}$& $0.38_{-0.03}^{+0.03}$& $0.55_{-0.03}^{+0.03}$& $0.6_{-0.3}^{+0.4}$& $1.7_{-0.7}^{+1.2}$\\
NGC4535
& $24.5_{-3.6}^{+8.5}$& $4.6_{-1.0}^{+2.2}$& $8.9_{-1.1}^{+2.3}$& $239_{-40}^{+60}$& $13.5_{-3.8}^{+3.0}$& $3.0_{-1.3}^{+3.1}$& $0.43_{-0.03}^{+0.03}$& $0.41_{-0.04}^{+0.04}$& $0.8_{-0.3}^{+0.6}$& $2.3_{-1.0}^{+1.7}$\\
NGC4540
& $14.0_{-2.9}^{+2.7}$& $<\,2.6$& $<\,7.0$& $<\,215$& $22.4_{-3.8}^{+12.6}$& $2.0_{-0.9}^{+1.5}$& $0.55_{-0.03}^{+0.04}$& $0.66_{-0.03}^{+0.03}$& $0.7_{-0.3}^{+0.5}$& $2.4_{-1.0}^{+1.7}$\\
NGC4548
& $13.9_{-2.6}^{+4.4}$& $<\,2.7$& $<\,6.9$& $<\,250$& $39.3_{-15.2}^{+17.1}$& $1.6_{-0.7}^{+1.4}$& $0.47_{-0.02}^{+0.03}$& $0.74_{-0.01}^{+0.01}$& $0.9_{-0.4}^{+0.6}$& $2.5_{-1.0}^{+1.8}$\\
NGC4569
& $16.1_{-2.5}^{+3.9}$& $<\,2.2$& $<\,6.5$& $380_{-61}^{+83}$& $72.3_{-22.4}^{+33.9}$& $1.1_{-0.5}^{+0.9}$& $0.61_{-0.03}^{+0.04}$& $0.27_{-0.04}^{+0.04}$& $1.4_{-0.6}^{+1.0}$& $6.5_{-2.7}^{+4.7}$\\
NGC4571
& $19.3_{-2.8}^{+5.7}$& $4.5_{-1.0}^{+1.7}$& $8.8_{-1.1}^{+1.8}$& $255_{-36}^{+50}$& $13.1_{-3.4}^{+3.2}$& $3.3_{-1.4}^{+2.8}$& $0.64_{-0.02}^{+0.02}$& $0.29_{-0.06}^{+0.07}$& $0.6_{-0.2}^{+0.4}$& $2.8_{-1.2}^{+2.0}$\\
NGC4654
& $19.8_{-3.7}^{+4.3}$& $4.9_{-1.5}^{+1.8}$& $9.1_{-1.5}^{+1.8}$& $441_{-79}^{+145}$& $23.4_{-4.9}^{+8.0}$& $2.8_{-1.2}^{+2.2}$& $0.34_{-0.04}^{+0.04}$& $0.45_{-0.04}^{+0.04}$& $0.7_{-0.3}^{+0.5}$& $2.0_{-0.9}^{+1.5}$\\
NGC4689
& $23.6_{-3.7}^{+4.1}$& $3.8_{-1.0}^{+1.0}$& $8.1_{-1.0}^{+1.0}$& $189_{-22}^{+25}$& $13.0_{-2.4}^{+4.0}$& $4.6_{-2.0}^{+3.5}$& $0.53_{-0.02}^{+0.02}$& $0.61_{-0.03}^{+0.03}$& $0.5_{-0.2}^{+0.4}$& $1.7_{-0.7}^{+1.2}$\\
NGC4731
& $13.1_{-2.8}^{+3.1}$& $2.6_{-1.0}^{+1.1}$& $7.0_{-1.1}^{+1.2}$& $248_{-49}^{+93}$& $24.8_{-6.5}^{+14.1}$& $1.8_{-0.8}^{+1.4}$& $0.36_{-0.05}^{+0.06}$& $0.44_{-0.07}^{+0.09}$& $0.7_{-0.3}^{+0.5}$& $1.6_{-0.6}^{+1.1}$\\
NGC4781
& $8.3_{-1.1}^{+1.5}$& $2.1_{-0.6}^{+0.8}$& $6.4_{-0.6}^{+0.9}$& $182_{-32}^{+53}$& $22.7_{-5.2}^{+6.4}$& $1.4_{-0.6}^{+1.1}$& $0.46_{-0.03}^{+0.04}$& $0.72_{-0.02}^{+0.02}$& $0.6_{-0.2}^{+0.4}$& $1.9_{-0.8}^{+1.4}$\\
NGC4941
& $21.1_{-3.6}^{+4.2}$& $2.7_{-0.9}^{+1.0}$& $7.0_{-1.0}^{+1.0}$& $273_{-36}^{+56}$& $23.1_{-5.8}^{+11.0}$& $2.5_{-1.1}^{+1.9}$& $0.57_{-0.02}^{+0.02}$& $0.87_{-0.01}^{+0.01}$& $0.8_{-0.4}^{+0.6}$& $3.2_{-1.3}^{+2.3}$\\
NGC4951
& $7.9_{-2.1}^{+4.4}$& $1.9_{-0.8}^{+1.6}$& $6.2_{-0.9}^{+1.7}$& $329_{-66}^{+130}$& $38.3_{-16.8}^{+27.2}$& $1.1_{-0.5}^{+1.2}$& $0.67_{-0.02}^{+0.02}$& $0.73_{-0.03}^{+0.05}$& $0.7_{-0.3}^{+0.5}$& $3.2_{-1.3}^{+2.3}$\\
NGC5042
& $14.7_{-2.8}^{+3.7}$& $2.5_{-1.0}^{+1.1}$& $6.8_{-1.0}^{+1.2}$& $242_{-45}^{+68}$& $23.2_{-6.8}^{+12.8}$& $2.0_{-0.9}^{+1.6}$& $0.48_{-0.02}^{+0.02}$& $0.65_{-0.03}^{+0.04}$& $0.7_{-0.3}^{+0.5}$& $2.0_{-0.8}^{+1.4}$\\
NGC5068
& $11.4_{-1.9}^{+2.3}$& $1.1_{-0.3}^{+0.3}$& $5.5_{-0.3}^{+0.4}$& $117_{-12}^{+18}$& $16.1_{-3.6}^{+5.3}$& $2.7_{-1.2}^{+2.1}$& $0.54_{-0.01}^{+0.01}$& $0.11_{-0.10}^{+0.14}$& $0.4_{-0.2}^{+0.3}$& $1.1_{-0.5}^{+0.8}$\\
NGC5134
& $15.0_{-2.9}^{+3.3}$& $2.5_{-0.9}^{+0.7}$& $6.8_{-0.9}^{+0.7}$& $229_{-37}^{+49}$& $21.2_{-4.1}^{+9.8}$& $3.8_{-1.6}^{+3.0}$& $0.49_{-0.02}^{+0.01}$& $0.67_{-0.03}^{+0.04}$& $0.4_{-0.2}^{+0.3}$& $1.2_{-0.5}^{+0.8}$\\
NGC5248
& $15.0_{-2.6}^{+3.3}$& $2.5_{-0.9}^{+1.1}$& $6.7_{-0.9}^{+1.2}$& $173_{-27}^{+40}$& $20.8_{-5.6}^{+8.8}$& $2.6_{-1.1}^{+2.1}$& $0.56_{-0.03}^{+0.03}$& $0.73_{-0.02}^{+0.02}$& $0.6_{-0.2}^{+0.4}$& $2.1_{-0.9}^{+1.5}$\\
NGC5530
& $20.4_{-2.8}^{+5.0}$& $4.3_{-0.7}^{+1.2}$& $8.6_{-0.8}^{+1.3}$& $269_{-33}^{+49}$& $11.6_{-2.4}^{+2.1}$& $2.3_{-0.9}^{+1.9}$& $0.51_{-0.02}^{+0.02}$& $0.70_{-0.02}^{+0.02}$& $0.9_{-0.4}^{+0.6}$& $2.7_{-1.1}^{+2.0}$\\
NGC5643
& $17.7_{-2.7}^{+2.7}$& $3.0_{-0.8}^{+0.7}$& $7.3_{-0.8}^{+0.8}$& $188_{-21}^{+33}$& $13.8_{-2.3}^{+4.4}$& $4.6_{-2.0}^{+3.5}$& $0.40_{-0.02}^{+0.02}$& $0.70_{-0.01}^{+0.01}$& $0.4_{-0.2}^{+0.3}$& $1.1_{-0.5}^{+0.8}$\\
NGC6300
& $21.5_{-2.9}^{+3.6}$& $3.6_{-0.8}^{+1.0}$& $7.8_{-0.9}^{+1.1}$& $156_{-18}^{+23}$& $11.0_{-2.3}^{+2.7}$& $3.2_{-1.4}^{+2.4}$& $0.49_{-0.02}^{+0.02}$& $0.67_{-0.02}^{+0.02}$& $0.7_{-0.3}^{+0.5}$& $2.0_{-0.8}^{+1.4}$\\
NGC6744
& $31.8_{-3.0}^{+3.7}$& $3.9_{-0.5}^{+0.7}$& $8.2_{-0.6}^{+0.8}$& $156_{-9}^{+13}$& $9.4_{-1.3}^{+1.3}$& $2.6_{-1.1}^{+1.9}$& $0.38_{-0.01}^{+0.01}$& $0.17_{-0.03}^{+0.03}$& $1.2_{-0.5}^{+0.9}$& $3.0_{-1.2}^{+2.1}$\\
NGC7456
& $11.0_{-2.3}^{+4.6}$& $2.6_{-0.9}^{+1.5}$& $7.1_{-1.0}^{+1.6}$& $393_{-80}^{+101}$& $34.7_{-12.3}^{+15.9}$& $0.8_{-0.3}^{+0.7}$& $0.36_{-0.05}^{+0.06}$& $0.81_{-0.02}^{+0.02}$& $1.4_{-0.6}^{+1.0}$& $3.5_{-1.5}^{+2.5}$\\
NGC7496
& $18.4_{-3.0}^{+10.5}$& $4.1_{-1.2}^{+3.7}$& $8.4_{-1.3}^{+3.7}$& $262_{-58}^{+161}$& $17.4_{-7.3}^{+5.5}$& $5.2_{-2.1}^{+6.3}$& $0.54_{-0.02}^{+0.01}$& $0.69_{-0.03}^{+0.04}$& $0.4_{-0.1}^{+0.3}$& $1.1_{-0.5}^{+0.8}$\\
\hline
\end{tabular}
\end{threeparttable}
\end{center}
\end{table*}

\subsection{Cloud lifetime ($t_{\rm CO}$)}\label{sssec:tco}
Across all the galaxies in our sample, the range of measured cloud lifetimes (i.e., the duration over which CO is visible) is 5\,${-}$30\,Myr, with an average of 16\,Myr and a $16{-}84$\% range of $11{-}22$\,Myr. The range of our measurements of $t_{\rm CO}$ corresponds to $1{-}3$ times the average crossing time-scale of massive GMCs in PHANGS--ALMA (see also J.~Sun et al. subm.; and Section \ref{subsec:dissc_model}), which suggests that clouds are transient objects that disperse within a small multiple of the dynamical time-scale. 

\begin{figure}
\includegraphics[scale=0.52]{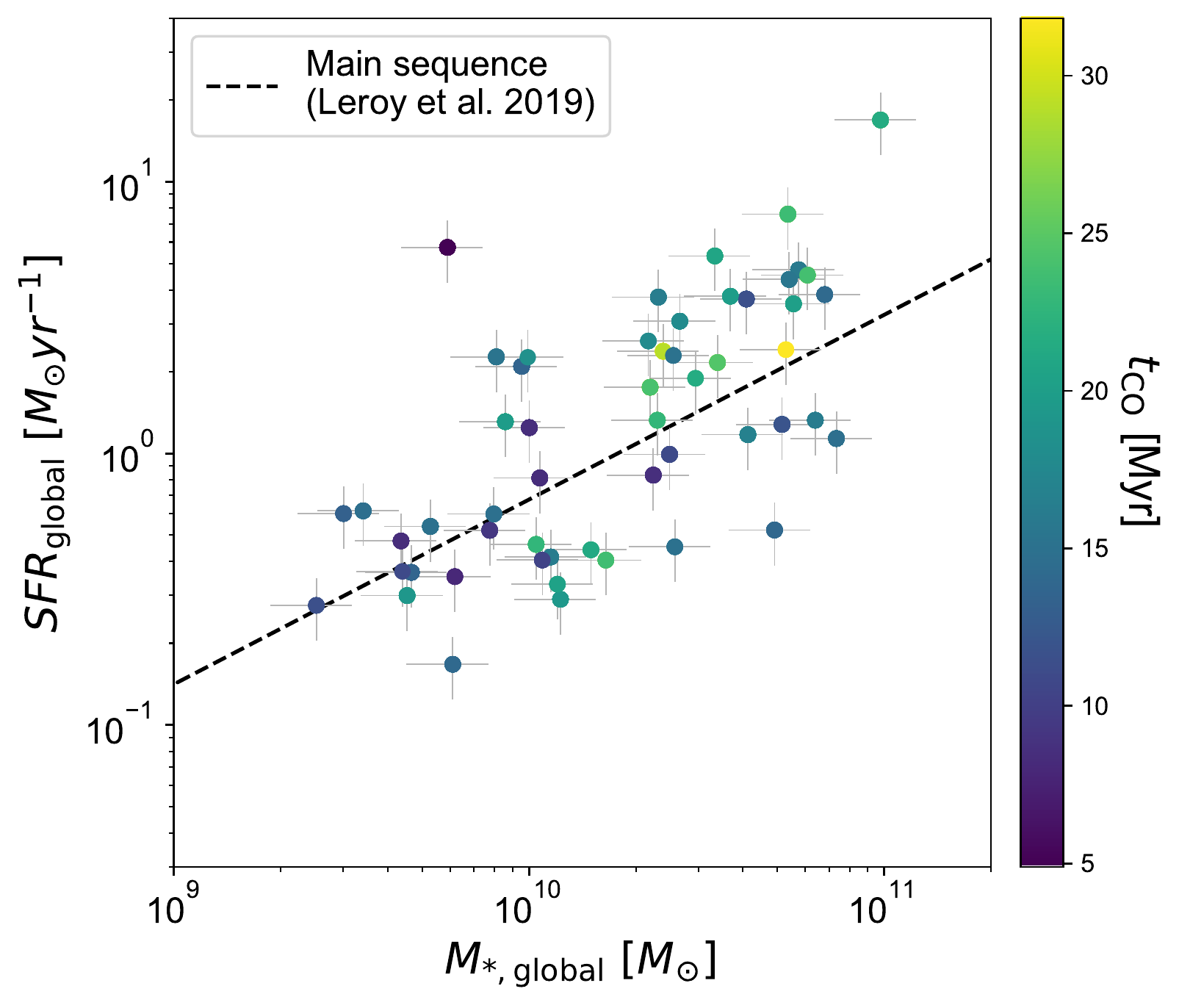}
\caption{Galaxy-wide SFR ($\rm SFR_{global}$) as a function of stellar mass ($M_{\rm *, global}$) for our full galaxy sample, colour-coded by our measurement of the cloud lifetime ($t_{\rm CO}$) for each galaxy. The dashed line is the local star-forming main sequence from \citet{leroy19}. } \label{fig:ms_tco}
\end{figure}

The overall measured range of molecular cloud lifetimes is consistent with that found in previous studies, those using cloud classification methods \citep{engargiola03, blitz07,fukui08, kawamura09, miura12, meidt15, corbelli17}, statistics of sight line fractions with only CO or only H$\alpha$ or both types of emission \citep{schinnerer19, pan22}, and those using the same statistical method as described in Section~\ref{sec:method} \citep{kruijssen19, chevance20, hygatePhD, kim21, ward22}. Similar cloud survival times have been predicted by theory and simulations \citep[e.g.][]{elmegreen00, hartmann01,dobbs13, kim18,benincasa20,jeffreson21, lancaster21, semenov21}. In Figure~\ref{fig:ms_tco}, we show how the measured cloud lifetime is correlated with the position of galaxies in the $M_{\rm *, global}-\rm SFR_{\rm global}$ plane. The figure shows that $t_{\rm CO}$ increases with $\rm SFR_{\rm global}$ and $M_{\rm *, global}$ (see also top left panel of Figure~\ref{fig:exsig}). There seems to be no relation between $t_{\rm CO}$ and the host galaxy's offset from the main sequence ($\rm\Delta MS$). \citet{pan22} have found similar trends for the PHANGS galaxies, in which the fraction of sight lines per galaxy that is associated only with CO emission increases with $M_{\rm *, global}$, while showing no correlation with $\rm\Delta MS$. Although not directly comparable, other time-scales such as the average free-fall time-scale measured on cloud scales and the global depletion time-scale have also been reported to correlate strongly with galaxy mass across the PHANGS sample \citep{utomo18}. The galaxy mass trend of $t_{\rm CO}$ is discussed in more detail in Section~\ref{subsec:results_envgal}.

\subsection{Feedback time-scale ($t_{\rm fb}$)}\label{sssec:tfb}
The duration over which CO and H$\rm\alpha$ emission is found coincident is measured to be less than $6.1$\,Myr in our sample of galaxies. For 12 galaxies, we do not sufficiently resolve the separation between independent regions and therefore we are only able to obtain upper limits on $t_{\rm fb}$ (see Appendix~\ref{sec:accuracy}). Without these galaxies, the range of feedback time-scale becomes $1.3{-}5.4$\,Myr, constituting $10{-}30\%$ of the cloud lifetime, with an average and a standard deviation of $3.2\pm1.1$\,Myr. This time-scale represents the time it takes for emerging massive stars (visible in H$\rm\alpha$) to disperse the surrounding molecular gas (either by kinetic dispersal or by the photodissociation of CO molecules). The range of feedback time-scales measured across our galaxy sample is comparable to that from our previous studies using the same statistical method \citep{kruijssen19, chevance20_fb, hygatePhD, kim21, ward22}. 

Our measurements of the feedback time-scales are also similar to the time it takes for optically identified stellar clusters and associations to stop being associated with their natal GMCs ($1-5$\,Myr; \citealp{whitmore14, hollyhead15, corbelli17, turner22}). \citet{grasha18, grasha19} have measured similar ages of star clusters and associations when they become spatially decorrelated from GMCs in NGC\,7793 and M51 (2 and 6\,Myr, respectively). Hydrodynamical simulations of GMCs \citep{kim18, kim20, grudic21, lancaster21} find somewhat longer feedback time-scales ($\lesssim 10$\,Myr), constituting $\sim50\%$ of the cloud lifetime. We suspect this difference could be due to different approaches for tracing star formation in simulations and observations. Indeed, simulations trace star formation by employing sink particles, which are created when a certain density threshold is reached assuming a fully populated initial mass function and include a phase of deeply embedded star formation. On the other hand, we focus on H$\rm\alpha$, which is sensitive to the most massive stars; in the case where the star formation accelerates over time (e.g., \citealp{hartmann12, murray15}), our measurements may be the most sensitive to the final, intense phase of star formation. Moreover, H$\rm\alpha$ is attenuated during the earliest phase of star formation due to the dense gas surrounding the young stars. Including 24~$\mu$m as a tracer for the obscured star formation increases the overlapping time-scale between CO and SFR tracer by $1{-}4$\,Myr \citep{kim21}.

\subsection{Region separation length ($\lambda$)}\label{sssec:lambda}
Figure~\ref{fig:tuningforks} reveals that there is a universal spatial decorrelation between molecular gas and young stellar regions on small spatial scales, while these quantities are correlated with each other on galactic scales. This result demonstrates that galaxies are composed of small regions undergoing independent evolution from GMCs to cold gas-free young stellar regions. Our method constrains the characteristic separation length ($\lambda$) between the small-scale independent regions, which is linked to the scale at which molecular gas-to-SFR tracer flux ratio starts to deviate from the galactic average (see Figure~\ref{fig:tuningforks}). Excluding 8 galaxies for which we do not sufficiently resolve these independent regions (see Appendix~\ref{sec:accuracy}), we find that $\lambda$ ranges from 100\,pc to 400\,pc, with an average and standard deviation of $254\pm 79$\,pc. This is similar to the total cold gas disc thickness (100-300\,pc; \citealp{scoville93, yim14, heyer15, patra20, yim20}), as well as the range of values found in previous application of the same method to relatively nearby and well-resolved galaxies ($100{-}250$~pc, \citealp{kruijssen19, chevance20, kim21}). From the similarity of $\lambda$ to the gas disc scale height, \citet{kruijssen19} have suggested that the break-out of feedback-driven bubbles from the galactic disc, pushing the ISM by a similar distance, might be setting this characteristic length scale.

While our methodology constrains the mean separation length between regions undergoing independent lifecycles, other methods focus on characterising the separation between detectable emission peaks. In a parallel paper on the PHANGS galaxies, Machado et al. (in prep.) investigate the spacing between emission peaks in the PHANGS CO maps. Contrary to our study, which uses the highest available resolution for each galaxy, they adopt GMC catalogue (A. Hughes in prep.; see also \citealp{rosolowsky21}) that are generated using CO maps with matched resolution of 150\,pc and sensitivity across the full sample. For a sub-set of 44 galaxies in our sample, Machado et al. (in prep.) obtain mean distances to the first nearest neighbour from 250\,pc to 600\,pc. We have compared these distances to the nearest neighbour distance expected from the mean separation length between GMCs, obtained by $\langle r_{\rm n}\rangle=0.443\lambda_{\rm GMC}=0.443\lambda\sqrt{\tau/t_{\rm CO}}$ \citep[see the discussion of][eq.~9]{kruijssen19}, where $\tau$ is the total duration of the entire evolutionary cycle ($\tau=t_{\rm CO}+t_{\rm H\alpha}-t_{\rm fb}$). The $\langle r_{\rm n}\rangle$ ranges from 50\,pc to 200\,pc. While the two quantities show a mild correlation (with Spearman correlation coefficient of 0.5), the mean distance to the first nearest neighbour from Machado et al. (in prep.) is larger than that expected from the mean separation length. We suspect this difference is due to the limitation in resolution (by 60\% coarser on average) of CO maps in Machado et al. (in prep.) compared to the maps analysed here, which results in a smaller number of identified GMCs compared to when high-resolution maps are used.

\subsection{Feedback velocity ($v_{\rm fb}$)}\label{sssec:vfb}
After the onset of star formation, the CO emission quickly becomes undetectable due to energetic feedback from young massive stars. We use the Gaussian 1$\sigma$ dispersion needed to reproduce the density contrast between the CO peaks and the local background ($r_{\rm CO}$) and the time-scale over which molecular clouds are disrupted ($t_{\rm fb}$) to define the feedback velocity $v_{\rm fb}=r_{\rm CO}/t_{\rm fb}$ (see also \citealp{kruijssen18}). The measured $v_{\rm fb}$ represents the speed with which the region must be swept free of CO molecules. The measurement does not specify a physical mechanism, but the most likely candidates are kinetic removal of gas from the region, e.g., by gas pressure-driven expansion, radiation pressure, winds, or supernovae, or the photodissociation of CO molecules by massive stars \citep[e.g.][]{barnes21, barnes22}. 

Excluding galaxies with resolution worse than 200\,pc (as $r_{\rm CO}$ depends on the beam size; see below), the size of the clouds is between 20 to 100\,pc and $v_{\rm fb}$ ranges between $10-50$\,\kms, with an average and standard deviation of $22\pm11$\,\kms. These measured cloud sizes are comparable to the luminosity-weighted averages of those derived from GMC catalogues for each galaxy (\citealp{rosolowsky21}, A.~Hughes in prep.), with Spearman correlation coefficient of 0.7. The range of velocities is consistent with that obtained from our previous analysis \citep{kruijssen19, chevance20, kim21} and is comparable to the expansion velocities measured for nearby H\,{\sc ii} regions in NGC300 \citep{mcleod20}, the LMC \citep{naze01,ward16,mcleod19}, and the Milky Way \citep{murray10, barnes20}. A similar range of expansion velocities is also found in numerical simulations by \citet{rahner17} and  \citet{kim18}.

We note that the measured $r_{\rm CO}$ depends on the beam size of the CO maps. If the CO emission peaks are dispersed kinetically, then the measured feedback velocity should be considered as accurate, because the measured $r_{\rm CO}$ is the same size scale over which the material must travel to achieve the spatial displacement necessary to cease the spatial overlap between CO and H$\rm\alpha$ emission. However, if the CO emission peaks are dispersed by photodissociation, then this spatial overlap may cease before the feedback front reaches $r_{\rm CO}$. In that case, $v_{\rm fb}$ may be subject to beam dilution and should be considered as an upper limit to the velocity of the dissociation front. For a sub-set of 19 galaxies with PHANGS--MUSE \citep{emsellem22} observations, \citet{kreckel20} and \citet{williams22} have measured the metallicity distribution, as well as the scale at which the mixing in the ISM is effective, using a two-point correlation function. \citet{kreckel20} have found a strong correlation between the mixing scale and $v_{\rm fb}$ (Pearson's correlation coefficient of 0.7), indicating that dispersal of molecular gas is predominantly kinetic. 

\begin{figure}
\includegraphics[scale=0.52]{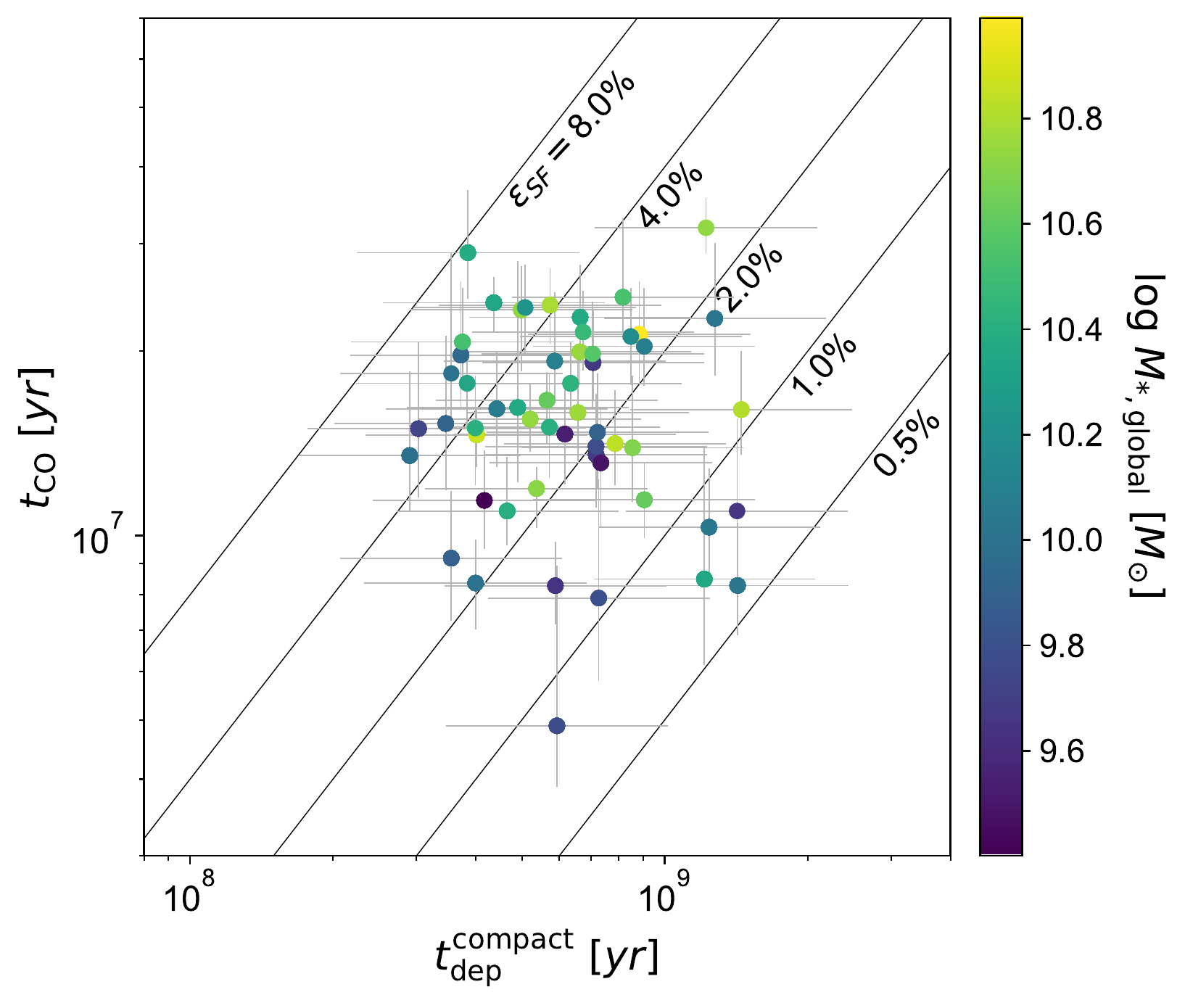}
\caption{Our measurements of cloud lifetime ($t_{\rm CO}$) against the depletion time for compact molecular gas ($t_{\rm dep}^{\rm compact}$). The data points are colour-coded by stellar mass ($M_{\rm *, global}$). Lines of constant $\epsilon_{\rm sf}$ are shown using equation~\ref{eq:esf}.  } \label{fig:tcotdep}
\end{figure}

\subsection{Integrated star formation efficiency ($\epsilon_{\rm sf}$)}\label{sssec:esf}
We define the integrated star formation efficiency per cloud lifecycle ($\epsilon_{\rm sf}$) as \begin{equation}
\label{eq:esf}
\epsilon_{\rm sf} = \frac{t_{\rm CO} \Sigma_{\rm SFR}}{\Sigma_{\rm H_{2}}^{\rm compact}}~,
\end{equation}
where $\Sigma_{\rm SFR}$ and $\Sigma_{\rm H_{2}}^{\rm compact}$ are the surface densities of SFR and compact molecular gas of the analysed region, respectively. This allows us to directly compare the rate of SFR ($\Sigma_{\rm SFR}$) and the rate at which molecular gas participating in the star formation enters and leaves molecular clouds, which can be expressed as $\Sigma_{\rm H_{2}}^{\rm compact }/t_{\rm CO}$. Equation~\ref{eq:esf} can also simply be rewritten as $\epsilon_{\rm sf}=t_{\rm CO}/t_{\rm dep}^{\rm compact}$, where $t_{\rm dep}^{\rm compact}=\Sigma_{\rm H_{2}}^{\rm compact}/\Sigma_{\rm SFR}$ is the depletion time-scale of non-diffuse molecular gas structures (clouds), assuming that all star formation takes place in such structures. When measuring $\Sigma_{\rm H_{2}}^{\rm compact}$, we take the sum of the compact CO emission and divide by the analysed area after filtering out diffuse emission. This is to selectively include CO emission that participates in massive star formation, while excluding CO emission from diffuse gas and faint clouds. However, to calculate $\Sigma_{\rm SFR}$, we include all the emission, assuming that all the diffuse emission in the SFR tracer map (WISE W4 band in combination with GALEX ultraviolet band; see Section~\ref{sec:method}) is related to recent massive star formation (e.g., leakage of ionising photons from H\,{\sc ii} regions). This assumption might not hold especially in the central region of galaxies where contributions from hot low-mass evolved stars in diffuse ionised gas is found to be non-negligible \citep{belfiore22}. However, galactic centres are mostly not included in our analysis.

Our measurements of $t_{\rm dep}^{\rm compact}$ are listed in Table~\ref{tab:result} and range from $0.3{-}1.4$\,Gyr. Since we only take the compact gas emission into account, $t_{\rm dep}^{\rm compact}$ is shorter than that measured including all the CO emission ($t_{\rm dep}$), which ranges from $0.9{-}7.6$\,Gyr for the fields of view considered here. The depletion time-scales of the PHANGS--ALMA galaxies across the full ALMA footprints (including galactic centres) can be found in \citet{utomo18}, \citet{leroy21_survey}, \citet{querejeta21}, and J.~Sun et al. (subm.). In Figure~\ref{fig:tcotdep}, we show the molecular cloud lifetime $t_{\rm CO}$ as a function of the gas depletion time of the compact molecular gas $t_{\rm dep}^{\rm compact}$, as measured in our analysis. Following the above procedure, we measure $\epsilon_{\rm sf}$ to be $0.7-7.5\%$ across our galaxy sample, illustrating that star formation is inefficient in these clouds. Our previous measurements of $\epsilon_{\rm sf}$ \citep{kruijssen19, chevance20, kim21} also fall within this range of values. 

We have also compared our measurements of $\epsilon_{\rm sf}$ with the star formation efficiency per free-fall time, defined as $\epsilon_{\rm ff}=t_{\rm ff}\Sigma_{\rm SFR}/\Sigma_{\rm H_{2}}$. Using a subset of PHANGS-ALMA observations, \citet{utomo18} have measured $\epsilon_{\rm ff}$ ($0.3-2.6\%$) that are similar to $\epsilon_{\rm sf}$ within a factor of few, where the difference is mostly because $t_{\rm CO}$ is on average 3 times longer than $t_{\rm ff}$ (see Section~\ref{subsec:dissc_model}).

\subsection{Diffuse emission fraction in CO and H$\alpha$ maps ($f_{\rm diffuse}^ {\rm CO}$ and $f_{\rm diffuse}^ {\rm  H\alpha}$)}\label{sssec:diffuse}

As described in Section~\ref{sec:method}, in order to robustly perform our measurements, we filter out the large-scale diffuse emission with a Gaussian high-pass filter in Fourier space. With this procedure, we can also constrain the fraction of emission coming from the diffuse component in both CO and H$\rm\alpha$ maps ($f_{\rm diffuse}^ {\rm CO}$ and $f_{\rm diffuse}^ {\rm  H\alpha}$, respectively). As shown in Table~\ref{tab:result}, we measure a fraction of diffuse CO emission ranging from 13 to 69\%, with an average of 45\%. We obtain diffuse ionised gas fractions ranging from 11 to 87\%, with an average of 0.53\%. We note that these values are determined directly from the morphological structure of the integrated emission maps. They do not contain any information regarding the dynamical state of the gas, nor do they account for galaxy-to-galaxy variations in resolution, sensitivity, or inclination. As a result, these diffuse emission fractions represent important functional quantities, but their physical interpretation may be non-trivial. \citet{pety13} have suggested a similar value of diffuse CO emission fraction in M51, finding that 50\% of the CO emission arises from spatial scale larger than 1.3\,kpc. \citet{roman-duval16} have measured 25\% of the CO emission in the Milky Way to be diffuse. As for the diffuse H$\alpha$ emission fraction, our range of values matches well with what is found in dedicated diffuse ionised gas studies based on H\,{\sc ii} region morphologies, where  \cite{belfiore22} have found the diffuse emission fraction to range from $20-55$\%, with a median of 37\%, for the galaxies in the PHANGS--MUSE sample. Using an un-sharp masking technique, \citet{pan22} have estimated $40-90$\% for the H$\alpha$ diffuse emission fraction for the galaxies in our sample. \citet{tomivic21} also finds a similar range of values for the diffuse ionised gas fraction in 70 local cluster galaxies.

\begin{table*}
\begin{center}
\caption{Spearman correlation coefficients and logarithm of $p$-values (in parentheses) between our measurements and galaxy (or average GMC) related properties. Our measurements are the cloud lifetime ($t_{\rm CO}$), the feedback time-scale ($t_{\rm fb}$), the characteristic region separation length ($\lambda$),the integrated star formation efficiency (\esf), the feedback velocity (\vfb), the diffuse emission fraction in the molecular gas and SFR tracer maps ($f_{\rm diffuse}^ {\rm CO}$ and $f_{\rm diffuse}^ {\rm  H\alpha}$), and the depletion time-scale of the compact molecular gas ($t_{\rm dep}^{\rm compact}$). Statistically significant correlations are in red (also in bold).} \label{tab:corr}
\begin{threeparttable}
\setlength\tabcolsep{2pt}
\begin{tabular}{lccccccccc}
\hline
 &$t_{\rm CO}$& $t_{\rm fb}$& $\lambda$ & $\epsilon_{\rm sf}$&$v_{\rm fb}$ & $f_{\rm diffuse}^ {\rm CO}$& $f_{\rm diffuse}^ {\rm  H\alpha}$& $t_{\rm dep}^{\rm compact}$\\
\hline
\multicolumn{4}{l}{Galaxy related properties}\\
\hline
Stellar mass, $M_{\rm *, global}$& \textcolor{red}{\textbf{0.47 (-3.33)}}& 0.34 (-1.58)& 0.13 (-0.43)& 0.04 (-0.10)& -0.02 (-0.04)& -0.13 (-0.42)& -0.17 (-0.61)& 0.31 (-1.60)\\
Galaxy-wide SFR, $\rm SFR_{global}$ &0.40 (-2.48)& 0.34 (-1.62)& 0.20 (-0.76)& 0.18 (-0.71)& 0.05 (-0.12)& -0.41 (-2.41)& -0.16 (-0.57)& 0.01 (-0.02)\\
Atomic gas mass, $M_{\rm HI, global}$ &0.12 (-0.42)& 0.08 (-0.21)& 0.12 (-0.37)& 0.00 (-0.01)& 0.14 (-0.42)& \textcolor{red}{\textbf{-0.52 (-3.83)}}& -0.28 (-1.36)& 0.02 (-0.05)\\
Molecular gas mass, $M_{\rm H_{2}, global}$ &\textcolor{red}{\textbf{0.59 (-5.33)}}& \textcolor{red}{\textbf{0.49 (-3.01)}}& 0.14 (-0.44)& 0.22 (-0.91)& -0.14 (-0.43)& -0.21 (-0.82)& -0.19 (-0.75)& 0.15 (-0.54)\\
Offset from the main sequence, $\rm\Delta MS$ &0.16 (-0.59)& 0.09 (-0.25)& 0.02 (-0.05)& 0.23 (-0.99)& 0.08 (-0.21)& -0.40 (-2.37)& -0.08 (-0.23)& -0.18 (-0.66)\\
Hubble type&-0.23 (-0.98)& -0.03 (-0.07)& -0.25 (-1.02)& -0.03 (-0.09)& -0.22 (-0.79)& -0.13 (-0.43)& -0.19 (-0.72)& -0.15 (-0.54)\\
Total gas mass, $M_{\rm gas, global}$&0.33 (-1.78)& 0.27 (-1.08)& 0.09 (-0.26)& 0.09 (-0.27)& 0.00 (-0.00)& \textcolor{red}{\textbf{-0.49 (-3.50)}}& -0.29 (-1.40)& 0.08 (-0.26)\\
Total baryonic mass, $M_{\rm tot, global}$&\textcolor{red}{\textbf{0.46 (-3.16)}}& 0.33 (-1.53)& 0.11 (-0.34)& 0.04 (-0.10)& -0.03 (-0.06)& -0.21 (-0.84)& -0.18 (-0.69)& 0.30 (-1.48)\\
Molecular gas fraction, $f_{\rm H_{2}, global}$  &\textcolor{red}{\textbf{0.50 (-3.72)}}& \textcolor{red}{\textbf{0.50 (-3.23)}}& 0.16 (-0.53)& 0.15 (-0.52)& -0.16 (-0.50)& 0.22 (-0.92)& 0.07 (-0.22)& 0.19 (-0.75)\\
Gas fraction, $f_{\rm gas, global}$ &-0.23 (-0.98)& -0.17 (-0.56)& -0.15 (-0.48)& 0.07 (-0.20)& -0.01 (-0.03)& -0.40 (-2.32)& -0.19 (-0.73)& -0.30 (-1.48)\\
Specific SFR, $\rm sSFR$& 0.07 (-0.22)& 0.08 (-0.20)& 0.05 (-0.14)& -0.23 (-0.98)& -0.08 (-0.22)& 0.33 (-1.71)& 0.02 (-0.06)& 0.36 (-2.09)\\
Metallicity, $\rm 12+log(O/H)$ &0.54 (-2.43)& 0.32 (-0.85)& 0.17 (-0.38)& -0.27 (-0.77)& 0.07 (-0.12)& -0.22 (-0.51)& 0.06 (-0.11)& \textcolor{red}{\textbf{0.66 (-3.77)}}\\
Mixing scale, $L_{\rm mix}$&-0.36 (-0.75)& 0.07 (-0.08)& \textcolor{red}{\textbf{0.78 (-3.05)}}& -0.02 (-0.03)& 0.66 (-1.85)& 0.10 (-0.14)& 0.52 (-1.40)& -0.30 (-0.60)\\
\hline
\multicolumn{4}{l}{Average GMC related properties}\\
\hline
Velocity dispersion, $\sigma_{\rm v, GMC}$&0.19 (-0.76)& 0.22 (-0.82)& 0.32 (-1.49)& 0.02 (-0.06)& 0.22 (-0.79)& -0.06 (-0.17)& 0.25 (-1.11)& 0.11 (-0.35)\\
Virial parameter, $\alpha_{\rm vir, GMC}$& -0.32 (-1.67)& -0.23 (-0.86)& -0.21 (-0.77)& -0.13 (-0.46)& -0.07 (-0.19)& -0.03 (-0.09)& \textcolor{red}{\textbf{0.47 (-3.26)}}& 0.06 (-0.16)\\
Molecular gas mass, $M_{\rm GMC}$ &0.37 (-2.12)& 0.38 (-1.94)& 0.40 (-2.26)& 0.22 (-0.93)& 0.15 (-0.46)& -0.03 (-0.08)& 0.05 (-0.14)& -0.04 (-0.09)\\
Internal pressure, $P_{\rm int}$  & 0.14 (-0.47)& 0.12 (-0.36)& 0.21 (-0.80)& 0.07 (-0.21)& 0.19 (-0.63)& -0.02 (-0.05)& 0.00 (-0.01)& -0.04 (-0.10)\\
Molecular gas surface density, $\Sigma_{\rm H_{2}, GMC}$&0.20 (-0.81)& 0.11 (-0.33)& 0.16 (-0.55)& 0.13 (-0.43)& 0.15 (-0.48)& 0.03 (-0.08)& -0.17 (-0.62)& -0.08 (-0.25)\\
\hline
\multicolumn{4}{l}{Galactic dynamics related properties}\\
\hline
Angular speed, $\Omega$ &-0.06 (-0.14)& 0.04 (-0.08)& -0.41 (-1.86)& -0.13 (-0.39)& -0.34 (-1.27)& 0.17 (-0.53)& 0.27 (-1.03)& 0.21 (-0.73)\\
Toomre stability parameter, $Q$  &-0.37 (-1.74)& -0.24 (-0.74)& 0.20 (-0.62)& -0.06 (-0.16)& 0.38 (-1.48)& -0.07 (-0.17)& 0.24 (-0.88)& -0.18 (-0.60)\\
Velocity dispersion, $\sigma$&0.11 (-0.34)& 0.35 (-1.45)& 0.23 (-0.77)& 0.10 (-0.29)& -0.02 (-0.05)& -0.24 (-0.92)& 0.30 (-1.37)& -0.04 (-0.09)\\
\hline
\multicolumn{4}{l}{Other derived quantities within our method}\\
\hline
Surface density\\
... molecular gas, ${\Sigma}_{\rm H_{2}}$ &\textcolor{red}{\textbf{0.44 (-3.02)}}& 0.41 (-2.21)& -0.07 (-0.19)& 0.05 (-0.14)& -0.23 (-0.83)& -0.02 (-0.06)& 0.00 (-0.01)& 0.24 (-1.03)\\
... compact molecular gas, $\Sigma_{\rm H_{2}}^{\rm compact}$ &\textcolor{red}{\textbf{0.43 (-2.81)}}& 0.40 (-2.13)& -0.11 (-0.32)& 0.07 (-0.20)& -0.25 (-0.96)& -0.16 (-0.58)& -0.04 (-0.09)& 0.21 (-0.90)\\
Total mass\\
... molecular gas, $M_{\rm H_{2}}$&\textcolor{red}{\textbf{0.54 (-4.37)}}& \textcolor{red}{\textbf{0.55 (-3.85)}}& 0.15 (-0.49)& 0.15 (-0.54)& -0.19 (-0.66)& -0.13 (-0.43)& -0.04 (-0.12)& 0.18 (-0.69)\\
... compact molecular gas, $M_{\rm H_{2}}^{\rm compact}$&\textcolor{red}{\textbf{0.53 (-4.25)}}& \textcolor{red}{\textbf{0.55 (-3.90)}}& 0.11 (-0.34)& 0.16 (-0.58)& -0.19 (-0.68)& -0.28 (-1.27)& -0.07 (-0.21)& 0.17 (-0.62)\\
SFR surface density, $\Sigma_{\rm SFR}$&0.28 (-1.36)& 0.34 (-1.59)& -0.12 (-0.39)& \textcolor{red}{\textbf{0.43 (-2.85)}}& -0.27 (-1.13)& -0.19 (-0.72)& -0.03 (-0.08)& -0.30 (-1.54)\\
SFR&\textcolor{red}{\textbf{0.47 (-3.29)}}& \textcolor{red}{\textbf{0.53 (-3.66)}}& 0.14 (-0.45)& 0.39 (-2.35)& -0.17 (-0.56)& -0.30 (-1.41)& -0.15 (-0.53)& -0.16 (-0.60)\\
CO emission density contrast, ${\cal E}_{\rm CO}$&-0.35 (-1.95)& \textcolor{red}{\textbf{-0.59 (-4.58)}}& -0.43 (-2.55)& -0.04 (-0.10)& -0.00 (-0.01)& 0.11 (-0.35)& -0.16 (-0.61)& -0.19 (-0.74)\\
H$\alpha$ emission density contrast, ${\cal E}_{\rm H\alpha}$&0.10 (-0.30)& -0.36 (-1.78)& -0.36 (-1.84)& 0.18 (-0.71)& -0.22 (-0.80)& 0.08 (-0.25)& -0.22 (-0.89)& -0.11 (-0.38)\\
\hline
\multicolumn{4}{l}{Observational systematic parameters}\\
\hline
Inclination, $i$&-0.26 (-1.17)& -0.26 (-1.01)& 0.20 (-0.73)& \textcolor{red}{\textbf{-0.57 (-4.90)}}& 0.40 (-2.14)& 0.07 (-0.20)& 0.35 (-1.93)& 0.40 (-2.54)\\
Resolution, $l_{\rm ap, min}$&0.04 (-0.11)& 0.30 (-1.32)& \textcolor{red}{\textbf{0.90 (-16.72)}}& -0.06 (-0.19)& \textcolor{red}{\textbf{0.54 (-3.67)}}& 0.07 (-0.18)& 0.28 (-1.37)& 0.01 (-0.03)\\
Noise&0.13 (-0.42)& -0.16 (-0.48)& -0.32 (-1.35)& -0.03 (-0.08)& -0.16 (-0.48)& 0.19 (-0.69)& \textcolor{red}{\textbf{-0.48 (-3.07)}}& 0.13 (-0.44)\\
Completeness&0.36 (-1.87)& 0.41 (-1.94)& 0.10 (-0.27)& 0.20 (-0.75)& -0.15 (-0.42)& -0.03 (-0.07)& 0.21 (-0.80)& 0.06 (-0.16)\\
\hline
\end{tabular}
\end{threeparttable}
\end{center}
\end{table*}

\section{Discussion}\label{sec:discussion}

\begin{table*}
\begin{center}
\caption{Summary of the 20 statistically significant and physically meaningful correlations identified in Table~\ref{tab:corr} between our measurements listed in Table~\ref{tab:result} and galaxy and average cloud properties. For each correlation, the table lists the Spearman correlation coefficient ($r$), the associated $p$-value, the slope of the best-fitting linear regression ${\rm d}y/{\rm d}x$, the intercept of the best fit $y_{\rm 0}$, and the scatter around the best-fitting relation. } \label{tab:corr_sum}
\begin{threeparttable}
\begin{tabular}{ccccccccc}
\hline
Quantity (y) & [units]& Correlates with(x)& [units] & Spearman $r$ & $\rm log$ Spearman $p$ & $\rm dy/dx$ & $y_{\rm 0}$ & Scatter  \\
\hline
$\rm log$\,$t_{\rm CO}$& [Myr] &$\rm log$\,$M_{\rm H_{2}, global}$ &[$M_{\rm \odot}$] &0.59&-5.33&0.16&-0.24&0.14\\
$\rm log$\,$t_{\rm CO}$& [Myr]&$\rm log$\,$M_{\rm H_{2}}$&[$M_{\rm \odot}$] &0.54&-4.37&0.16&-0.22&0.14\\
$\rm log$\,$t_{\rm CO}$& [Myr]&$\rm log$\,$M_{\rm H_{2}}^{\rm compact}$&[$M_{\rm \odot}$]&0.53&-4.25&0.17&-0.22&0.14\\
$\rm log$\,$t_{\rm CO}$& [Myr]&$\rm log$\,$f_{\rm H_{2}, global}$ & [-] &0.50&-3.72&0.23&1.30&0.14\\
$\rm log$\,$t_{\rm CO}$&[Myr] &$\rm log$\,$M_{\rm *, global}$& [$M_{\rm \odot}$] &0.47&-3.33&0.16&-0.40&0.14\\
$\rm log$\,$t_{\rm CO}$& [Myr]&$\rm log$\,$\rm SFR$& [$M_{\rm \odot}\rm yr^{-1}$] & 0.47&-3.29&0.14&1.24&0.14\\
$\rm log$\,$t_{\rm CO}$ & [Myr]&$\rm log$\,$M_{\rm tot, global}$& [$M_{\rm \odot}$] &0.46&-3.16&0.17&-0.50&0.14\\
$\rm log$\,$t_{\rm CO}$& [Myr]&$\rm log$\, $\Sigma_{\rm H_{2}}$& [$M_{\rm \odot}\rm pc^{-2}$] &0.44&-3.02&0.17&1.02&0.15\\
$\rm log$\,$t_{\rm CO}$&[Myr] &$\rm log$\, $\Sigma_{\rm H_{2}}^{\rm compact}$& [$M_{\rm \odot}\rm pc^{-2}$] &0.43&-2.81&0.19&1.10&0.15\\
\hline
$\rm log$\,$t_{\rm fb}$ &[Myr] &$\rm log$\,${\cal E}_{\rm CO}$ & [-]&-0.59&-4.58&-0.61&1.02&0.13\\
$\rm log$\,$t_{\rm fb}$ &[Myr] &$\rm log$\, $M_{\rm H_{2}}^{\rm compact}$& [$M_{\rm \odot}$]&0.55&-3.90&0.24&-1.56&0.15\\
$\rm log$\,$t_{\rm fb}$ &[Myr] &$\rm log$\,$M_{\rm H_{2}}$& [$M_{\rm \odot}$] &0.55&-3.85&0.23&-1.64&0.15\\
$\rm log$\,$t_{\rm fb}$ & [Myr]&$\rm log$\,$\rm SFR$& [$M_{\rm \odot}\rm yr^{-1}$]&0.53&-3.66&0.24&0.53&0.15\\
$\rm log$\,$t_{\rm fb}$ & [Myr]&$\rm log$\,$f_{\rm H_{2}, global}$ & [-] &0.50&-3.23&0.31&0.60&0.16\\
$\rm log$\,$t_{\rm fb}$ &[Myr] &$\rm log$\,$M_{\rm H_{2}, global}$ &[$M_{\rm \odot}$] &0.49&-3.01&0.20&-1.33&0.16\\
\hline
$\rm log$\,$\lambda$&[pc]&$\rm log$\,$L_{\rm mix}$& [pc] &0.78&-3.05&0.84&0.18&0.08\\
\hline
$\rm log$\,$\epsilon_{\rm sf}$&[-] &$\rm log$\,$\Sigma_{\rm SFR}$&[$M_{\rm \odot}\rm yr^{-1}pc^{-2}$]&0.43&-2.85&0.31&-1.79&0.19\\
\hline
$f_{\rm diffuse}^{\rm CO}$&[-] &$\rm log$\,$M_{\rm HI, global}$&[$M_{\rm \odot}$]   &-0.52&-3.83&-0.13&1.63&0.11\\
$f_{\rm diffuse}^{\rm CO}$&[-]  &$\rm log$\,$M_{\rm tot, global}$& [$ M_{\rm\odot}$] &-0.49&-3.50&-0.13&1.67&0.11\\
\hline
$f_{\rm diffuse}^{\rm H\alpha}$&[-]  &$\rm log$\,$\rm\alpha_{vir, GMC}$& [-]  &0.47&-3.26&0.30&0.48&0.17\\
\hline
\end{tabular}
\end{threeparttable}
\end{center}
\end{table*}

\subsection{Relations with global galaxy and average cloud properties}\label{subsec:results_envgal}

We have correlated our measurements shown in Table~\ref{tab:result} with global properties of galaxies and luminosity-weighted average properties of the cloud population in each galaxy. The properties considered are listed in Table~\ref{tab:corr}. We use the galaxy properties listed in Table~\ref{tab:sample}, as well as combinations of these quantities. We derive the total baryonic mass of the galaxy ($M_{\rm tot, global}=M_{\rm *, global}+M_{\rm HI, global}+M_{\rm H_{2}, global}$), total gas mass ($M_{\rm gas, global}=M_{\rm HI, global}+M_{\rm H_{2}, global}$),  molecular gas fraction ($f_{\rm H_{2}, global}= M_{\rm H_{2}, global}/M_{\rm gas, global}$), gas fraction ($f_{\rm gas, global}=M_{\rm gas, global}/ M_{\rm tot, global}$), and specific SFR (sSFR\,$=\rm SFR_{global}/$ $M_{\rm *, global}$). We also look for correlations with gas phase metallicity [$\rm 12+log(O/H)$] for the subset of 23 galaxies for which direct measurements are available. These measurements are taken from \citet{kreckel19} for the 18 galaxies in our sample with MUSE observations and from \citet{pilyugin14} for 5 galaxies that do not overlap with the MUSE sample. We also include the 50\% correlation scale ($L_{\rm mix}$) of the two-dimensional metallicity distribution maps (after metallicity gradient subtraction) of the PHANGS--MUSE sample from \citet{kreckel20} and \citet{williams22}. This scale indicates the length over which the mixing in the ISM is effective, and ranges from 200 to 600\,pc.

The luminosity-weighted averages of the cloud properties are determined from the GMC catalogues that have been established for the full PHANGS--ALMA sample using the CPROPS algorithm (\citealp{rosolowsky21}, Hughes et al.\ in prep ). Here, we use measurements of the cloud velocity dispersion ($\sigma_{\rm v, GMC}$), virial parameter $\alpha_{\rm vir, GMC}$, mass ($M_{\rm GMC}$), internal pressure ($P_{\rm int}$), and molecular gas surface density ($\Sigma_{\rm H_{2}, GMC}$).

Metrics related to galactic dynamics are included using measurements of the rotation curve ($v_{c}$) as a function of radius ($r$) from \citet{lang20}. These metrics are the angular speed ($ \Omega=v_{\rm c}(r)/r$) and the Toomre stability parameter of the mid-plane molecular gas ($\rm Q= \kappa {\sigma}_{H_{2}}/\pi G \Sigma_{H_{2}}$), where ${\sigma}_{\rm H_{2}}$ is the velocity dispersion measured from CO moment 2 maps, at native resolution, and $\kappa$ $=\Omega\sqrt{2(1+\beta)}$ with $ \beta=\rm d\,ln$\,$v_{\rm c}(R)/\rm d\,ln$\,$R$, numerically calculated. Since these values vary with galactocentric radius, we first divide the galaxy into five different radial bins and calculate $\Omega$, $Q$, and ${\sigma}_{\rm H_{2}}$ for each bin. We then calculate the CO luminosity-weighted average of these values.

We explored possible correlations with galaxy global properties constrained within our method. These are different from the values listed in the Table~\ref{tab:sample} in that they are calculated within the analysed region, i.e.\ excluding galactic bulge and bar in most galaxies, and restricted to regions where CO observations have been made. Moreover, we provide two individual measurements for the molecular gas mass surface density and total molecular gas mass, where one takes only the compact emission into account (denoted as $\Sigma_{\rm H_{2}}^{\rm compact}$ and $M_{\rm H_{2}}^{\rm compact}$, which our measurements of the time-scales are based on) and the other includes all the emission (denoted as $\Sigma_{\rm H_{2}}$ and $ M_{\rm H_{2}}$). The quantities ${\cal E}_{\rm CO}$ and ${\cal E}_{\rm H\alpha}$ are the surface density contrast between the average emission of CO (respectively H$\rm\alpha$) peaks and the galactic average value, measured on the filtered map.
 
In order to explore possible systematic biases, we also include our minimum aperture size ($l_{\rm ap, min}$ in Table~\ref{tab:input}, which matches our working resolution), inclination ($i$; column (g) in Table~\ref{tab:sample}), completeness of CO observations, and noise of the CO data cube (in mK units) from \citet{leroy21_survey} as metrics. Finally we note that none of the properties listed here are corrected for galaxy inclination.

\begin{figure*}
\includegraphics[scale=0.72]{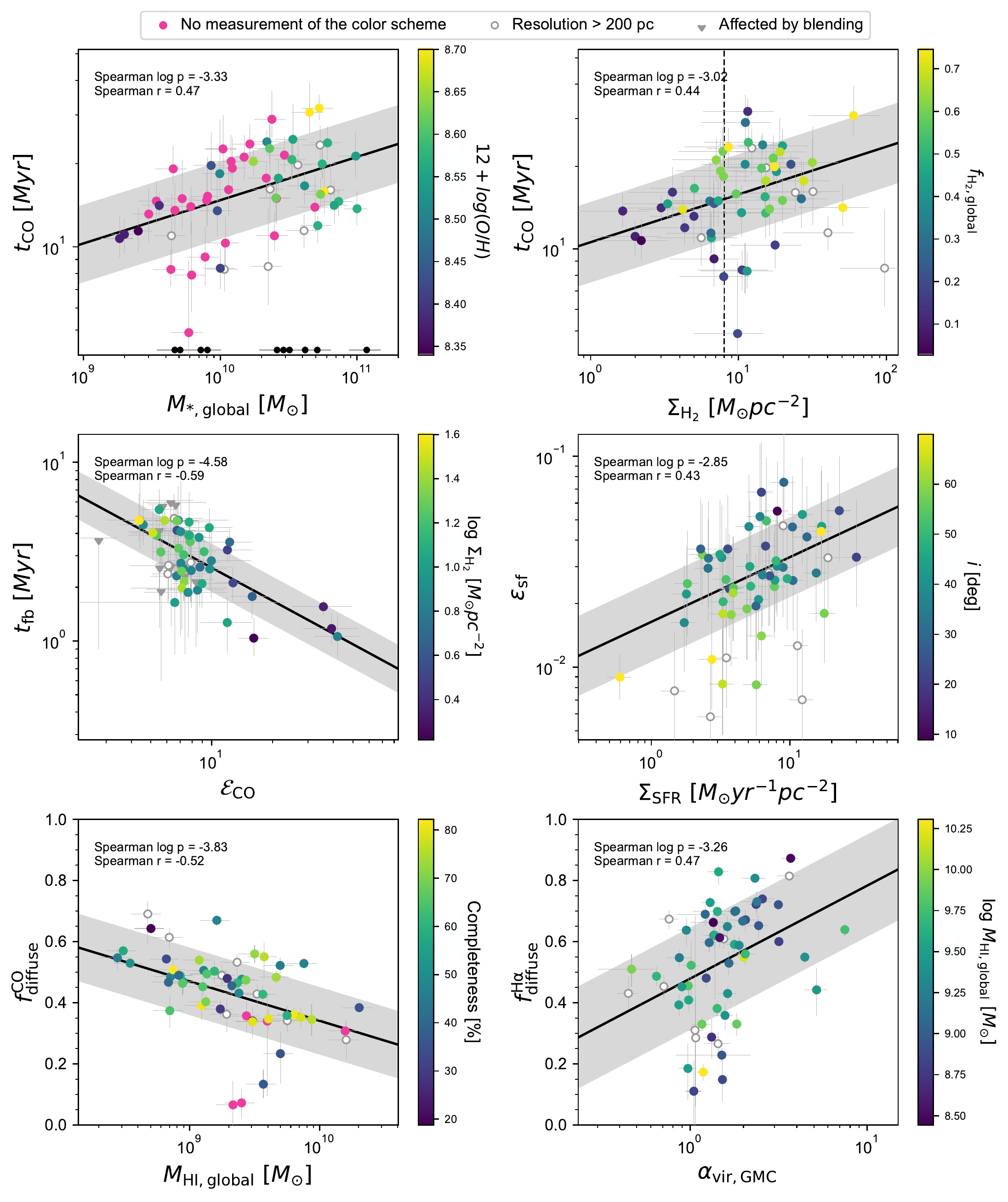}
\caption{Examples of six statistically significant correlations between our measurements and galaxy or average GMC properties. In the upper panel, the measured cloud lifetime ($t_{\rm CO}$) is shown as a function of stellar mass ($ M_{\rm *, global}$; left) and molecular gas surface density ($\Sigma_{\rm H_{2}}$; right), where points are colour coded by metallicity ($\rm 12+log\,(O/H)$) and molecular gas fraction ($f_{\rm H_{2}, global}$), respectively. The critical density identified by \citet{chevance20} is also shown for comparison (dashed line). In the middle left panel, the measured feedback time-scale ($t_{\rm fb}$) is shown as a function of the surface density contrast between CO emission peaks and the galactic average (${\cal E}_{\rm CO}$), colour coded by the molecular gas surface density ($\Sigma_{\rm H_{2}}$). The middle right panel shows the integrated star formation efficiency ($\epsilon_{\rm sf}$) as a function of SFR surface density ($\Sigma_{\rm SFR}$), where the points are colour coded by inclination ($i$). The lower left panel shows the diffuse CO emission fraction ($f_{\rm diffuse}^ {\rm CO}$) as a function of the global atomic gas mass ($M_{\rm HI, global}$), where the points are colour coded by completeness of the CO observation \citep{leroy21_survey}. In the lower right panel, the diffuse H$\alpha$ emission fraction ($f_{\rm diffuse}^ {\rm H\alpha}$) is shown as a function of average virial parameter of GMCs ($\alpha_{\rm vir, GMC}$), colour coded by the global atomic gas mass ($M_{\rm HI, global}$). Galaxies without measurements corresponding to each colour bar scheme are denoted in magenta. Gray circles are the galaxies with resolution worse than 200\,pc. Gray triangles indicate upper limits for galaxies suffering from blending of sources (see Section~\ref{sec:accuracy}). These gray points are excluded in our correlation analysis. For individual data points, 1$\sigma$ uncertainties are shown. In each panel, the best-fitting linear regression (solid line), 1$\sigma$ scatter of the data around the fit (shaded area), and the Spearman $p$-values and correlation coefficients are indicated. We also show the distribution of $M_{\rm *, global}$ for the ten galaxies excluded from our sample due to their limited number of emission peaks (black circles; see Section~\ref{sec:data} and Section~\ref{sec:accuracy}).} \label{fig:exsig}
\end{figure*}

\subsubsection{Statistically (in)significant correlations}\label{sssec:stat_corr}

In Table~\ref{tab:corr}, for all correlations, we list the Spearman rank correlation coefficients and the associated $p$-values, which represents the probability of a correlation appearing by chance. When evaluating the correlations, we exclude 8 galaxies where the resolution at which the analysis can be run is larger than 200\,pc, as we are likely to not sufficiently spatially separate star-forming regions in these galaxies (NGC1546, NGC1559,  NGC1672, NGC1792, NGC3511, NGC4569, NGC4654, NGC7456). Whenever a measurement of individual galaxies is considered as an upper limit, we also exclude the galaxy from our correlation analysis of a given measurement (see Appendix~\ref{sec:accuracy}). There are 12 galaxies (IC1954, NGC1087, NGC1097, NGC1385, NGC1546, NGC1672, NGC2090, NGC3627, NGC4298, NGC4540, NGC4548, NGC4569) with only upper limits of $t_{\rm fb}$ constrained. For 8 of these (IC1954, NGC1087, NGC1385, NGC1546, NGC1672, NGC4298, NGC4540, NGC4548) $\lambda$ is also an upper limit. Finally, we include six nearby galaxies (IC342, the LMC, M31, M33, M51, NGC300; previously analysed by \citet{kruijssen19}, \citet{chevance20}, and \citet{kim21}, which extend the range of environmental properties. 

We define a correlation to be statistically significant when the measured $p-$value is lower than $p_{\rm eff}$, where $p_{\rm eff}$ is derived using the Holm-Bonferroni method \citep[for an explanation and for an astrophysical application also see \citealt{kruijssen19b}]{holm79}. This method is used to account for the fact that spurious significant correlations may appear when comparisons between a large number of parameters are made. Specifically, we proceed by asking whether each of our measurement (columns in Table~\ref{tab:corr}) correlates with any of the galaxy and average cloud properties (rows in  Table~\ref{tab:corr}). We then rank the correlations by increasing $p$-value. For each correlation with a rank ($i$) of $i\rm \geq1$, we calculate the effective maximum $p$-value ($p_{\rm eff}$) below which the correlation is deemed significant (i.e.\ with $p<p_{\rm eff}$). We use the definition $p_{\rm eff}=p_{\rm ref}/(N_{\rm corr}+1-i)$, with $p_{\rm ref}\rm=0.05$ the desired confidence level and $N_{\rm corr}$ the number of independent variables being evaluated. In order to determine $N_{\rm corr}$, we subtract variables among the galaxy and average cloud properties that are trivially correlated. We find that numerous properties ($M_{\rm *, global}$, $M_{\rm HI, global}$, $M_{\rm H_{2}, global}$,  $\rm SFR_{global}$, $\rm \Delta\,MS$, $M_{\rm tot, global}$, $M_{\rm gas, global}$, $M_{\rm GMC}$, $P_{\rm int}$, $ \Sigma_{\rm H_{2}, GMC}$, $M_{\rm H_{2}}$, $M_{\rm H_{2}}^{\rm compact}$, and $\rm SFR$) are correlated significantly, with correlation coefficient higher than 0.7. We treat these parameters as one metric and this results in $N_{\rm corr}\approx20$. Table~\ref{tab:corr} shows statistically meaningful strong correlations in red, identified according to our definition. We note that even assuming that all the variables are independent does not significantly change our result as the $p$-values of strong correlations are all very small, with $\rm log$\,$(p)$ ranging from $-16.7$ to $-2.8$.

As shown in Table~\ref{tab:corr}, we identify biases of our measured quantities caused by the spatial resolution of maps ($l_{\rm ap, min}$) and inclination. Specifically, we find that (1) $l_{\rm ap, min}$ shows a strong correlation with $\lambda$ and $v_{\rm fb}$; (2) the inclination shows a strong correlation with $\epsilon_{\rm sf}$; (3) the noise of the CO data cube correlates with $f_{\rm diffuse}^{\rm H\alpha}$; and (4) the metallicity correlates with $t_{\rm dep}^{\rm compact}$. The covariance between resolution and $v_{\rm fb}=r_{\rm CO}/t_{\rm fb}$ is due to the increased measured cloud size ($r_{\rm CO}$) as the resolution gets worse. The dependence of $\lambda$ on $l_{\rm ap, min}$ implies that the measured region separation length ($\lambda$) would be biased upward when using maps that have poor resolution. However, despite the dependencies on resolution of these quantities, we are confident that the measured time-scales are less sensitive to the spatial resolution of the maps, because we require star-forming regions to be sufficiently resolved for our time-scale measurements to be considered as robust ($\lambda>1.5\,l_{\rm ap, min}$; see Section ~\ref{sec:accuracy} and \citealt{kruijssen18}). Unlike $\lambda$, the measured time-scales indeed do not show strong correlations with $l_{\rm ap, min}$. The dependence of $\epsilon_{\rm sf}=t_{\rm CO}/t_{\rm dep}^{\rm compact}$ on inclination is driven by a highly-significant correlation between inclination and $t_{\rm dep}^{\rm compact}$ ($\rm log$\,$(p)\rm=-2.5$; see Table~\ref{tab:corr}). We suspect that this latter correlation arises, because the filtering of the diffuse CO emission is less effective for highly inclined galaxies, and because the extinction correction applied to SFR maps may depend on inclination, as suggested by \citet{pellegrini20}. The dependence between the noise of the CO data cube and $f_{\rm diffuse}^{\rm H\alpha}$ seems to arise by a random chance, despite applying a strict threshold of $p$-values for correlations to be considered significant. Indeed, there is no logical link why these two quantities should show correlation and when the three galaxies with high noise level are excluded from the analysis, the strong correlation disappears. Finally, for the correlation between metallicity and $t_{\rm dep}^{\rm compact}$, we conjecture it could be related to the fact that the low-mass (low-metallicity) galaxies tend to have more diffuse emission due to their low surface brightness \citep{leroy21_survey}. The surface brightness sensitivity of the CO maps is not good enough to isolate the small clouds in low-mass (and low-metallicity) galaxies, which may therefore lead to more diffuse emission and low completeness for such galaxies \citep{leroy21_survey}. However, we note that completeness of the CO maps does not show a strong trend with $t_{\rm dep}^{\rm compact}$ (see Table~\ref{tab:corr}). The adopted metallicity-dependent $\rm \alpha_{CO}$ might also contribute to this observed trend between metallicity and $t_{\rm dep}^{\rm compact}$. While we partially correct for the presence of molecular gas that is not traced by CO emission (CO-dark gas) with this conversion factor, the observed strong correlation seems to indicate that the correction is insufficient. In closing, we again emphasise that our main measurements ($t_{\rm CO}, t_{\rm fb}$, and $\lambda$) are not affected by our prescription of $\rm \alpha_{CO}$, as they are based on relative changes of flux ratios (i.e.\ the global H$\rm\alpha$/CO ratio does not affect the time-scale estimate).

\subsubsection{Physical interpretation of significant correlations}\label{sssec:phys_corr}

In Table~\ref{tab:corr_sum}, we list the best-fitting relations using linear regressions, as well as their Spearman correlation coefficients and $p-$values, for statistically significant correlations in red in Table~\ref{tab:corr}, while correlations illustrating biases in our analysis (described in Section~\ref{sssec:stat_corr}), are excluded. Figure~\ref{fig:exsig} shows examples of six main strong correlations between our measurements and global galaxy properties. In this figure, we do not show all of the statistically meaningful correlations listed in Table~\ref{tab:corr_sum}, as they seem to be redundant and driven by correlations within galaxy properties (especially mass related quantities), and also within time-scales ($t_{\rm CO}$ and $t_{\rm fb}$). For example, as described above, $M_{\rm *, global}$ strongly correlate with $M_{\rm H_{2}, global}$, $M_{\rm H_{2}}$,  $M_{\rm H_{2}}^{\rm compact}$, $M_{\rm tot, global}$, $\rm SFR$, and $M_{\rm HI, global}$. By construction, $f_{\rm H_{2}, global}$ and $M_{\rm H_{2}, global}$ are not independent. Also, $\Sigma_{\rm H_{2}}$ and $\Sigma_{\rm H_{2}}^{\rm compact}$ correlate with each other. The correlation within time-scales is most likely due to the fact that our time-scale measurements are constrained by scaling the time-scale ratios with a reference time (see Section~\ref{sec:method}). 

Here, we offer explanations for how the relations in Table~\ref{tab:corr_sum} can be understood physically. However, we do not attempt to investigate which galaxy (or average GMC) properties are the main driver for these trends, because numerous properties also correlate with each other, making it hard to assess. 

First of all, the cloud lifetime ($t_{\rm CO}$) is measured to be longer with increasing stellar mass ($M_{\rm *, global}$), which traces galaxy mass. The cloud lifetime also shows positive correlations with the total molecular gas mass, both measured globally ($M_{\rm H_{2}, global}$) or only considering the analysed region, with and without diffuse emission ($M_{\rm H_{2}}$ and $M_{\rm H_{2}}^{\rm compact}$, respectively). Given that galaxy mass and metallicity are correlated (see Figure~\ref{fig:exsig}; upper left), we suspect this can be due to the fact that a higher gas density threshold is required to make CO visible in low mass galaxies compared to high mass galaxies. As shown in Table~\ref{tab:corr}, when only the galaxies with direct metallicity measurements are considered, a suggestive positive trend between $t_{\rm CO}$ and metallicity is revealed (Spearman correlation coefficient of 0.54), but this tentative trend is not strong enough to be characterised as statistically meaningful. CO molecules in low-metallicity environments with low dust-to-gas ratio are photodissociated deeper into the clouds \citep{bolatto13}. As clouds assemble from diffuse gas and become denser, clouds in a low-mass and low-metallicity environment spend a longer time in a CO-dark molecular gas phase (see also \citealp{clark12}). This is not included in the cloud lifetime we measure, because it is based on the visibility of CO emission, leading to an underestimation of the cloud lifetime. This is supported by the fact that, when HI emission is used to trace the gas, HI overdensities exist for a much longer duration prior to the formation of CO peaks \citep{ward20_HI}. High-mass galaxies also have a higher mid-plane pressure, which shapes clouds within the galaxy to have a higher internal pressure \citep{sun20b}, resulting in higher (surface) densities and thus making them easier to detect throughout their lifecycles \citep{wolfire10}.

Observational biases may also contribute to the relation between cloud lifetime and galaxy mass, where the CO emission in low-mass galaxies is typically lower than the noise level of the PHANGS--ALMA data. In low-mass galaxies, we might simply lack sensitivity to the CO emission to pick up emission from low mass GMCs at any point of their lifetime, while they are detected in high mass galaxies. \citet{schinnerer19} and  \citet{pan22} also find a higher fraction of CO-emitting sight lines in high-mass galaxies compared to low-mass galaxies and discuss that this trend is due to intrinsically low visibility of CO emission in low-mass galaxies. However, we expect this effect of sensitivity to be minor in our analysis, as our measurements are based on flux measurements and thus biased towards bright regions.

Other strong correlations with the cloud lifetime and properties related to the stellar mass and/or molecular gas mass of the galaxy (SFR and $M_{\rm tot, global}$) are likely to be driven by the correlations explained above. Several studies have also reported such connections between global properties of the galaxy and the ensemble average properties of clouds \citep{hughes13, colombo14,hirota18, sun18, sun20, sun20b, schruba19}. In the upper left panel of Figure~\ref{fig:exsig}, we also include the distribution of $M_{\rm *, global}$ for ten galaxies that are excluded from our sample due their small number of emission peaks (see Section~\ref{sec:data} and Section~\ref{sec:accuracy}). They are randomly distributed in $M_{\rm *, global}$, indicating that selectively including CO-bright galaxies does not bias this result.  

The cloud lifetime also positively correlates with the molecular gas fraction ($f_{\rm H_{2}, global}$), as well as with molecular gas surface densities measured with and without diffuse emission ($\Sigma_{\rm H_{2}}$ and $\Sigma_{\rm H_{2}}^{\rm compact}$, respectively). The relation with molecular gas surface density might seem to contradict theoretical expectations (e.g., \citealp{kim18}), because denser clouds are expected to collapse faster, form stars and disperse more quickly than lower-density clouds. As proposed by \citet{chevance20}, the observed strong correlation might be related to the transition from an atomic gas-dominated to a molecular gas-dominated environment, as shown by the coloured points in the upper-right panel of Figure~\ref{fig:exsig}. \citet{chevance20} have found that at a critical density threshold of $8~\msun\,\pc^{-2}$ (similar to the gas phase transition threshold), the cloud lifetime shows a better agreement with the galactic dynamical time-scale above this threshold and with the internal dynamical time-scale below. This value is similar to the molecular gas surface density at which the gas phase transition occurs $\sim\rm10\,M_{ \odot}\rm pc^{-2}$, at near solar-metallicity \citep[e.g.][]{wong02, bigiel08, leroy08, schruba11}. In the upper right panel of Figure~\ref{fig:exsig}, this transition density is shown as a dashed line for comparison. In an atomic gas-dominated environment ($f_{\rm H_{2}}<0.5$), CO is only emitted by the central region of the clouds, tracing the densest regions. However, in a molecular gas-dominated environment ($f_{\rm H_{2}}>0.5$), we detect more CO emission coming from an extended envelope of the molecular clouds \citep[e.g.][]{shetty14}. This may increase the measured cloud lifetimes, as the assembly phase of the envelope is additionally taken into account, compared to when only the densest phase is included. In addition, in a low-surface density environment, the clouds will spend a longer time in the CO-dark phase, as a higher density threshold is required to make CO visible, resulting in a measured cloud lifetime shorter than the actual molecular cloud assembly time \citep{bolatto13}. Similarly to the dependence on galaxy mass discussed above, observational biases due to CO sensitivity level also play a role, making us miss a higher fraction of low-mass clouds in atomic gas-dominated environments.

For the feedback time-scale, during which CO and H$\rm\alpha$ overlap, we find the strongest correlation with ${\cal E}_{\rm CO}$, which is the surface density contrast in the CO map between the emission peaks and the galactic average. The feedback time-scale becomes shorter with increasing ${\cal E}_{\rm CO}$. In the middle-left panel of Figure~\ref{fig:exsig}, points are colour coded by $\Sigma_{\rm H_{2}}$ and suggest that when ${\cal E}_{\rm CO}$ is higher (i.e. sharper CO emission peaks), feedback-driven dispersal of the clouds makes the CO emission become undetected faster. This can be understood physically as the CO emission will become invisible faster after the onset of star formation when the surrounding medium is sparse, indicated by the low molecular gas surface density, allowing a faster dispersal of molecular clouds. 

Similarly to the dependencies we have identified for $t_{\rm CO}$, we find that $t_{\rm fb}$ also correlates with $M_{\rm H_{2}}^{\rm compact}$, $M_{\rm H_{2}}$, $M_{\rm H_{2}, global}$, SFR, and $f_{\rm H_{2}, global}$. We suspect these correlations arise at least partially because $t_{\rm fb}$ and $t_{\rm CO}$ strongly correlate with each other (with a Spearman correlation coefficient of 0.72). Interestingly, unlike $t_{\rm CO}$, we find a less significant correlation with stellar mass, which does not satisfy our significance cut with $p\rm=0.03$. This can be explained by the fact that the feedback time captures the phase when clouds are star-forming, implying that the density is high enough, which typically corresponds to a CO-bright phase. Therefore, we miss less of the CO-dark phase that is proportionally more important in the low-mass galaxies. 

The mean separation length between independent regions ($\lambda$), which is linked to the scale at which molecular gas and young stars start to become spatially decorrelated, shows a strong positive correlation with the mixing scale traced by metallicity measurements of H\,{\sc ii} regions in PHANGS--MUSE galaxies from \citet{kreckel20} and \citet{williams22}. This trend indicates that for galaxies with broader and more efficient mixing in the ISM, molecular gas and young stars are separated by a larger distance. This might be physically understood, because a broader mixing length, most likely driven by stellar feedback, will push the gas further away from its original position. This would imply that the dispersal of the molecular cloud after star formation is kinetically driven, rather than by the photodissociation of the CO molecules. However, we note that this correlation could be at least partially driven by the resolution as it becomes weaker (with Spearman correlation coefficient from 0.78 to 0.66 with $p=0.02$) when the two galaxies with the highest resolution (NGC\,0628 and NGC5068) are excluded.

We find a strong correlation between our measurements of integrated star formation efficiency ($\epsilon_{\rm sf}$) and SFR surface density ($\Sigma_{\rm SFR}$), as shown in the middle right panel of Figure~\ref{fig:exsig}. This correlation might seem like it can be simply understood as that a higher integrated star formation efficiency per star formation event, at least, partially would be driven by a higher SFR. However, we cannot rule out the possibility that this relation arises due a strong correlation between $t_{\rm CO}$ and $\Sigma_{\rm H_{2}}^{\rm compact}$ explained above, making $\epsilon_{\rm sf}$ to be mostly dependent on $\Sigma_{\rm SFR}$ by construction (see Equation~\ref{eq:esf}). Moreover, potentially not enough extinction correction for highly inclined galaxies, shown as coloured points in Figure~\ref{fig:exsig}, can also contribute to this trend.

For the diffuse CO emission fraction ($f_{\rm diffuse}^{\rm CO}$), which is measured within our method during the diffuse emission filtering process of \citet{hygate19}, we find the strongest anti-correlation with the HI mass ($M_{\rm HI, global}$). Another correlation with $M_{\rm tot, gas}$ is driven by this strong correlation with $M_{\rm HI, global}$. We find that as $M_{\rm HI, global}$ decreases, the diffuse molecular component becomes more important. We conjecture that this can be due to an observational bias, as the completeness of the CO maps (indicated as coloured data points in Figure~\ref{fig:exsig}) is low for low-mass galaxies due to our limited sensitivity, suggesting that we are missing a larger fraction of small, faint clouds in such galaxies. However, contrary to our expectation, $f_{\rm diffuse}^{\rm CO}$ and completeness of CO emission maps from \citet{leroy21_survey} do not reveal a strong trend with each other (see Table~\ref{tab:corr}).  

Lastly, the diffuse H$\rm\alpha$ emission fraction ($f_{\rm diffuse}^{\rm H\alpha}$) shows a strong correlation with the average virial parameter of GMCs ($\alpha_{\rm vir, GMC}$). We suspect this can be due to a more pervasive medium of GMCs (higher $\alpha_{\rm vir, GMC}$) allowing more ionising photons to leak out from star-forming regions, compared to more bound clouds. Furthermore, as indicated by the coloured points in Figure~\ref{fig:exsig}, galaxies with higher $f_{\rm diffuse}^{\rm H\alpha}$ tend to have lower atomic gas mass ($M_{\rm HI, global}$), which is in line with observations showing deeper penetration of ionising photons into the surrounding interstellar medium in lower-mass (and lower-metallicity) galaxies \citep{cormier15,chevance16}.

\begin{figure*}
\includegraphics[scale=0.51]{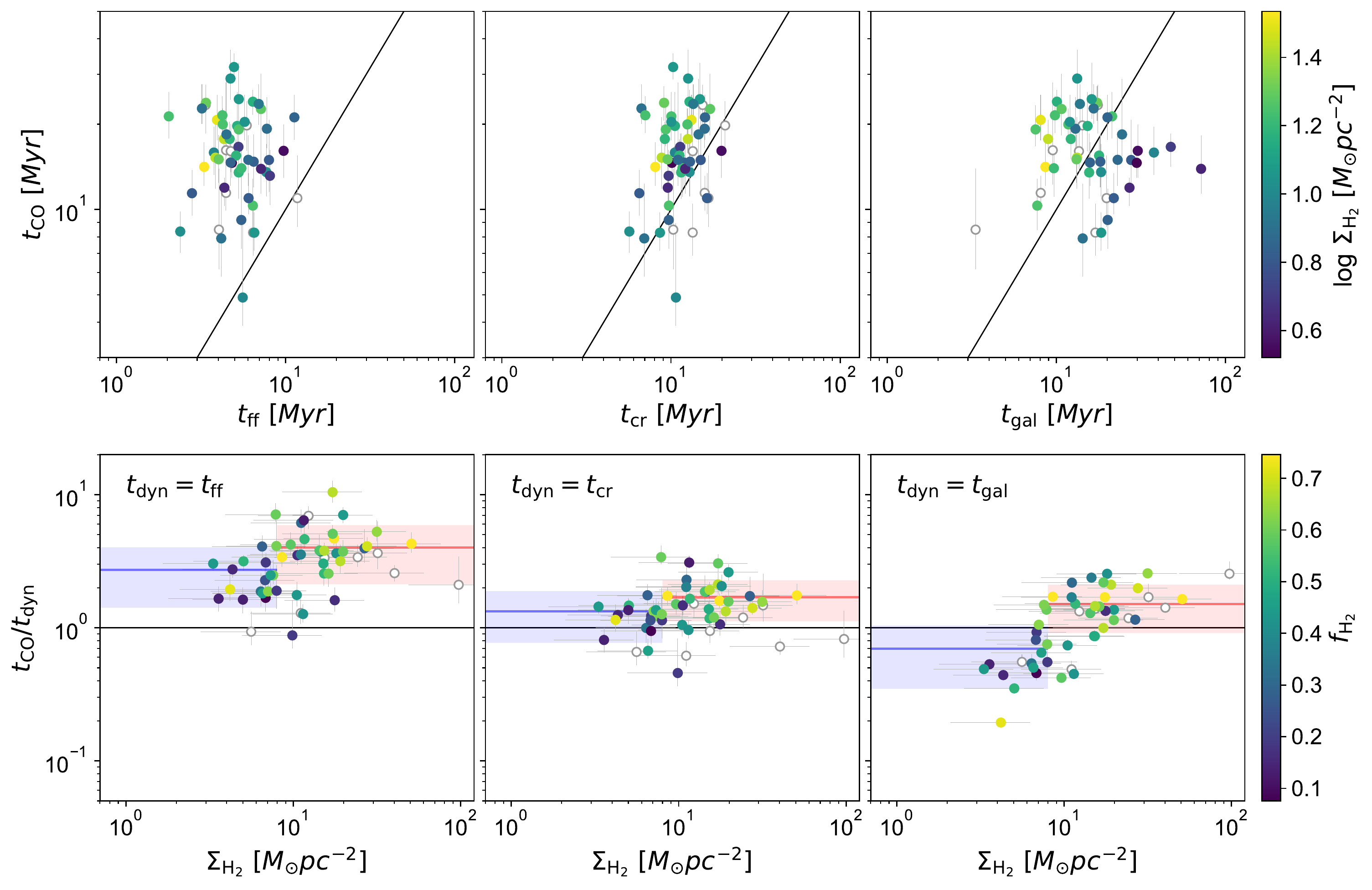}
\caption{Comparison of our measured cloud lifetime ($t_{\rm CO}$) with analytical predictions, which are, from left to right, the free-fall time ($t_{\rm ff}$), the crossing time of GMCs ($t_{\rm cr}$), and the galactic time-scale, considering the effect of large-scale dynamics ($t_{\rm gal}$). In the upper panels, the one-to-one relation is shown as a solid line and the data points are colour coded by surface molecular gas density ($\rm \Sigma_{H_{2}}$). The lower panels show the ratios of $t_{\rm CO}$ and analytical predictions ($t_{\rm dyn}$), where $t_{\rm dyn}$ is the free fall time (left), the crossing time (middle) or the galactic dynamical time-scale (right), as a function of $\rm \Sigma_{H_{2}}$, where the points are colour coded by molecular gas fraction ($f_{\rm H_{2}}$). In all panels, galaxies where our analysis can only be run at resolution larger than 200\,pc are shown as gray circles. The black horizontal line shows where $t_{\rm CO}=t_{\rm dyn}$. The blue and red lines, respectively, indicate the mean of the time-scale ratios for galaxies below and above $\rm \Sigma_{H_{2}}$ of 8$ M_{\rm\odot}\rm pc^{-2}$ (from \citealp{chevance20}) with the shaded regions representing the $\rm 16^{th}-84^{th}$ percentile. This density threshold is from \citet{chevance20}, below which $t_{\rm CO}$ shows better agreement with internal dynamical times ($t_{\rm ff}$ and $t_{\rm cr}$) compared to the $t_{\rm gal}$, and the other way around at densities higher than this threshold.} \label{fig:theory_time}
\end{figure*}

\subsection{Comparison with characteristic time-scales} \label{subsec:dissc_model}
In order to understand the dynamical mechanisms that govern cloud lifetimes, we compare our measurements of $t_{\rm CO}$ with analytical predictions, which are related to local cloud dynamics (GMC free-fall time and GMC crossing time; $t_{\rm ff}$ and $ t_{\rm cr}$, respectively), or large-scale dynamics of the ISM ($t_{\rm gal}$; \citealp{jeffreson18}). We adopt the CO-luminosity-weighted harmonic average of $\rm t_{ff}$ from the cloud catalogue established for the PHANGS--ALMA galaxies by A. Hughes (in prep.), which uses the CPROPS algorithm \citep{rosolowsky21} to determine the physical properties of GMCs. Here, $t_{\rm cr}$ is defined as $r_{\rm GMC}/\sigma_{\rm vel}$, where $ r_{\rm GMC}$ is the radius of GMC and $\sigma_{\rm vel}$ is the cloud velocity dispersion ($\sigma_{\rm vel}$). We first obtain $t_{\rm cr}$ for each GMC in a given galaxy and calculate CO-luminosity-weighted harmonic average of $t_{\rm cr}$. For $t_{\rm gal}$, we use the analytic theory presented in \citet{jeffreson18}. This theory assumes that cloud lifetimes are environmentally dependent and can be estimated by combining the time-scales of numerous processes governing the cloud evolution: the gravitational collapse of the mid-plane ISM ($\tau_{\rm ff}$), galactic shear ($\tau_{\beta}$), spiral arm interactions ($\tau_{\rm \Omega_{P}}$), epicyclic perturbations ($\tau_{\kappa}$), and cloud-cloud collisions ($\tau_{\rm cc}$). As the galactic shear ($\tau_{\beta}$) is a dynamically dispersive process unlike the other four mechanisms compressing the clouds, the cloud lifetime is expressed as $t_{\rm gal}^{-1}=|\tau_{\rm ff}+\tau_{\rm \Omega_{P}}+\tau_{\kappa}+\tau_{\rm cc}-\tau_{\beta}| $. We determine these time-scales in five different radial bins for each galaxy (see also Section~\ref{subsec:results_envgal}), using equations from \citet{jeffreson18}, the radial profiles of velocity dispersion from CO moment 2 maps, and the the rotational curves measured by \citet{lang20}. We then calculate harmonic averages of $t_{\rm gal}$ obtained in five radial bins. Five galaxies, for which the rotation curve is not available, are omitted from this comparison.

In the upper panel of Figure~\ref{fig:theory_time}, we show our measurements of $t_{\rm CO}$ as functions of the analytically predicted time-scales. In the lower panel, the ratios of $ t_{\rm CO}$ over these predicted time-scales are shown as a function of $\Sigma_{\rm H_{2}}$. For comparison, we also calculate the mean and 1$\sigma$ range of these ratios only considering galaxies with molecular gas surface density below and above the atomic-to-molecular transition density described in the previous subsection ($8\,M_{\odot}\,\rm pc^{-2}$; \citealt{chevance20}). We find that on average, $t_{\rm CO}=(2.7\pm1.3)\,t_{\rm ff}$ and $t_{\rm CO}=(4.0\pm1.9)\,t_{\rm ff}$, respectively, below and above the transition density with error bars representing 16-84\% range. Measured cloud lifetimes $t_{\rm CO}$ are almost always longer than $t_{\rm ff}$, independently of the molecular gas surface density. The crossing time $t_{\rm cr}$ shows a better agreement with $t_{\rm CO}$ compared to $t_{\rm ff}$, especially in low molecular gas surface density environments ($\Sigma_{\rm H_{2}}\leq 8\,\msun~\pc^{-2}$), with an average of $t_{\rm CO}=(1.3 \pm 0.6)\,t_{\rm cr}$. At higher densities, $ t_{\rm CO}$ increasingly deviates from $t_{\rm cr}$, with an average of $t_{\rm CO}=(1.7 \pm 0.6)\,t_{\rm cr}$. 

Finally, the right-hand panels of Figure~\ref{fig:theory_time} show that, in low molecular gas surface density environments, our measurements of $t_{\rm CO}$ are significantly lower than $t_{\rm gal}$, with an average of $t_{\rm CO}=(0.7 \pm 0.3)\,t_{\rm gal}$. By contrast, in higher-density ($\rm\geq 8\,M_{\odot}~\rm pc^{-2}$; molecular gas dominated) environments, where the majority of the molecular gas has high CO emissivity, $t_{\rm CO}$ becomes even longer than $t_{\rm gal}$ for most of the galaxies, with an average of $t_{\rm CO}=(1.5 \pm 0.6)\,t_{\rm gal}$. This difference in two density regimes can be explained by the fact that $t_{\rm gal}$ takes into account all the cold gas, including the phase that is not strongly CO-emitting, while our measurements of $t_{\rm CO}$ are based on CO-bright clouds. Therefore, $t_{\rm gal}$ is typically longer than $t_{\rm CO}$, especially in low molecular gas surface density environments. Moreover, this indicates that in low-surface density environments (atomic gas-dominated), the assembly of CO-bright molecular clouds seems to be less coupled to galactic dynamics, but rather occurs on internal cloud-dynamical time-scales such as the crossing time. Following the analytic theory by \citet{jeffreson18}, we can also quantify which galactic-dynamical mechanisms are relevant for setting the analytical cloud lifetime. The relevance is deemed significant when the time-scales of a given dynamical process is smaller than $2\times \tau_{\rm min}$, where $\tau_{\rm min}$ is the smallest time-scale among the five processes. The minimum time-scale has the greatest influence on setting the $t_{\rm gal}$, and is $\tau_{\rm ff}$ in most cases whereas shear is also found to be relevant in some environments with $\tau_{\beta}=(3.7\pm2.3)\,\tau_{\rm min}$, on average. This shows that the predicted lifetime of these molecular clouds results from a competition between the gravitational collapse of the mid-plane ISM and galactic shear, which causes clouds to be pulled apart by differential rotation. At high density where $t_{\rm CO}$ shows a good agreement with $t_{\rm gal}$, this implies that the assembly of molecular clouds and their evolution may be significantly influenced by galactic dynamics. \citet{meidt15} have reached a similar conclusion using GMCs in M51, where GMC evolution appears to be regulated by shear.

\section{Conclusion}\label{sec:concl}
We present a systematic determination of evolutionary sequences of GMCs from the molecular gas phase to exposed young stellar regions across an unprecedented sample of 54 molecular gas-rich main sequence galaxies from the PHANGS--ALMA survey \citep{leroy21_survey}. We have applied the statistical method developed by \citet{kruijssen14} and \citet{kruijssen18} to CO and H$\rm\alpha$ emission maps at cloud-scale resolution ($\sim$100\,pc) and measured the cloud lifetime ($t_{\rm CO}$), the feedback time-scale (duration for which CO and H$\rm\alpha$ are spatially overlapping; $t_{\rm fb}$), as well as the average separation length between independent star-forming regions evolving from molecular clouds to exposed young stellar regions ($\lambda$). We also derive other physical quantities such as the feedback velocity ($v_{\rm fb}$), the integrated star formation efficiency ($\epsilon_{\rm sf}$), and the diffuse emission fraction for both CO and H$\rm\alpha$ maps ($f_{\rm diffuse}^ {\rm CO}$ and $f_{\rm diffuse}^ {\rm  H\alpha}$). By capitalising on a statistically representative sample of galaxies from PHANGS, we have correlated our measurements with global galaxy and average cloud properties. This allows us to quantitatively link galactic-scale environmental properties to the small-scale evolutionary cycle of molecular clouds, star formation and feedback. The metrics explored here include properties related to galaxy mass, surface density of molecular gas and SFR, morphology, metallicity, velocity dispersion, pressure, and galactic dynamics. Our main conclusions are as follows:

\begin{enumerate} 
\item Across our sample of galaxies, we find that molecular clouds assemble and survive for a time-scale of $16.4\pm5.5$\,Myr on average, demonstrating that GMCs are transient objects that disperse after a few dynamical times via feedback from young massive stars. The feedback time-scale is $3.2\pm1.1$\,Myr on average  (excluding galaxies for which only an upper limit could be constrained) and constitutes $10-30\,\%$ of the cloud lifetime. Our measurements of these time-scales are in good agreement with those obtained using other methods (cloud classification based on their stellar content, e.g., \citealp{kawamura09,corbelli17}; determination of gas-free stellar cluster ages, e.g., \citealp{grasha19}). Our results further confirm the conclusion of previous works that there is a decorrelation between gas and young stars on the cloud scale \citep{schruba10,kreckel18,schinnerer19,pan22}, which can be explained by assuming that galaxies are composed of regions, undergoing evolution from gas to stars, that are separated by $100-400$\,pc on average. 

\item We find that the star formation in these regions is inefficient, with an integrated cloud-scale star formation efficiency ($\epsilon_{\rm sf}$) ranging from $0.8-7.5\%$. We measure feedback velocities ($v_{\rm fb}$) of $\rm 10-30\,km\,s^{-1}$. Overall, these results are consistent with those from our previous measurements, conducted on a significantly smaller number of galaxies \citep{kruijssen19, chevance20, kim21, chevance20_fb}. We also determine the fraction of diffuse emission in each CO and H$\rm\alpha$ map based on its morphology. We find average fractions of $45\pm10$\% in CO and $53\pm 19$\% in H$\alpha$. 

\item We find several statistically significant correlations between our measurements and global galaxy (or average cloud) properties (Table~\ref{tab:corr_sum}). In brief, $t_{\rm CO}$ shows positive correlations with quantities related to galaxy mass, as well as with the molecular gas surface density. These correlations can be explained by the existence of a CO-dark phase, during which the molecular clouds are beginning to assemble. Indeed, we miss more of this phase at low-mass (low-metallicity) and less dense environments as these environments require higher gas column density in order to shield CO molecules from being dissociated, compared to high-mass and high-density environment. Moreover, in high-surface density environments, we also capture the extended region of the GMCs, unlike in low-surface density (atomic gas-dominated) environments where CO is only tracing the densest centres of the GMCs. This results in longer cloud lifetimes in galaxies with a higher molecular gas surface density.

\item The feedback time-scale $t_{\rm fb}$ also shows correlations with quantities related to galaxy mass, most likely because $t_{\rm CO}$ and $t_{\rm fb}$ are correlated. However, $t_{\rm fb}$ does show an interesting relation with ${\cal E}_{\rm CO}$, which is the surface density contrast measured on a CO map between emission peaks and the galactic average. We find that $t_{\rm fb}$ is shorter with increasing ${\cal E}_{\rm CO}$ (i.e.\ towards sharper emission peaks). This can be physically understood as the result of feedback, where CO emission becomes undetected faster after the onset of massive star formation when the surrounding medium is more sparse. 

\item The star formation efficiency $\epsilon_{\rm sf}$ shows a strong correlation with $\Sigma_{\rm SFR}$, which can at least partly be understood as a higher SFR leading to a higher integrated star formation efficiency per star formation event. However, other factors can also contribute to this trend such as, the tight correlation between $t_{\rm CO}$ and $\Sigma_{\rm H_{2}}$, as well as the dependence of extinction correction on inclination.

\item We find a strong negative correlation with the diffuse gas fraction ($f_{\rm diffuse}^{\rm CO}$) and the global atomic gas mass ($M_{\rm HI, global}$). We attribute this correlation to the low completeness of CO observations in low-mass galaxies.

\item Diffuse H$\rm\alpha$ emission fraction strongly correlates with average virial parameter of GMCs ($\alpha_{\rm vir, GMC}$), which seems to indicate that a more pervasive medium of less bound GMCs allows more ionising photons to escape the star-forming region and to penetrate deeper into the surrounding gas. 

\item We find that, at all the density regimes probed here, $t_{\rm CO}$ is longer than $t_{\rm ff}$ (from \citealp{rosolowsky21}; A. Hughes in prep.) by a factor of $3.5\pm1.8$. By contrast, we find a good agreement with crossing time (from \citealp{rosolowsky21}; A. Hughes in prep.) with $t_{\rm CO}=(1.5\pm0.6)\,t_{\rm cr}$. The agreement becomes better when only the galaxies with low molecular gas surface density ($\Sigma_{\rm H_{2}}<8M_{\odot}\rm pc^{-2}$) are considered, with $t_{\rm CO}=(1.3\pm0.6)\,t_{\rm cr}$. At higher surface densities, the agreement becomes slightly worse with $t_{\rm CO}=(1.7\pm0.6)\,t_{\rm cr}$. Lastly, in the low-density regime ($<8~\msun~\pc^{-2}$), $t_{\rm CO}$ is shorter than the time-scale expected for galactic-dynamical processes to act, with $(0.7\pm0.3)\,t_{\rm gal}$, implying that $t_{\rm gal}$ overpredicts the cloud lifetime traced by CO emission. However, in higher surface density environments, $t_{\rm CO}=(1.5\pm0.6)\,t_{\rm gal}$, $t_{\rm CO}$ even becomes longer than $t_{\rm gal}$. The difference in low-surface density environments is likely due to the fact that GMCs spend a large fraction of their lifetime being CO-dark, and this phase is by construction excluded from $t_{\rm CO}$, which measures the CO visibility lifetime. By contrast, the \citet{jeffreson18} model does not make any distinction regarding on the CO emissivity of the different gas phases, and $t_{\rm gal}$ includes both the CO-dark and CO-bright phases. This results in an increase of $t_{\rm CO}/t_{\rm gal}$ with the molecular gas surface density \citep[also see][]{chevance20}.
\end{enumerate}

We have quantified the evolutionary lifecycle of GMC formation, evolution, and dispersal across an unprecedented sample of 54 nearby disc galaxies. We have demonstrated that this lifecycle depends on the large-scale galactic environment. In this work, we have determined the evolution from cold gas to exposed young stellar regions using CO and H$\rm\alpha$ maps. In the future, we plan to further extend and refine this evolutionary timeline, for a subset of our galaxy sample, by including other observations at different wavelengths: ionised emission lines from MUSE, mid-infrared from the \textit{James Webb Space Telescope}, and HI from the VLA and Meerkat. This will allow us to determine the time-scales of all the successive phases of the gas that participate in star formation.

\section*{Acknowledgements}
JK, MC, and JMDK gratefully acknowledge funding from the Deutsche Forschungsgemeinschaft (DFG, German Research Foundation)  through the DFG Sachbeihilfe (grant number KR4801/2-1).
MC gratefully acknowledges funding from the DFG through an Emmy Noether Research Group (grant number CH2137/1-1).
JMDK and MC gratefully acknowledge funding from the DFG through an Emmy Noether Research Group (grant number KR4801/1-1), as well as from the European Research Council (ERC) under the European Union's Horizon 2020 research and innovation programme via the ERC Starting Grant MUSTANG (grant agreement number 714907). 
EC acknowledges support from ANID BASAL projects ACE210002 and FB210003.
KG is supported by the Australian Research Council through the Discovery Early Career Researcher Award (DECRA) Fellowship DE220100766 funded by the Australian Government. K.G. is supported by the Australian Research Council Centre of Excellence for All Sky Astrophysics in 3 Dimensions (ASTRO~3D), through project number CE170100013. 
KK gratefully acknowledges funding from the German Research Foundation (DFG) in the form of an Emmy Noether Research Group (grant number KR4598/2-1, PI Kreckel). 
HAP acknowledges support by the Ministry of Science and Technology of Taiwan under grant 110-2112-M-032-020-MY3. 
MQ acknowledges support from the Spanish grant PID2019-106027GA-C44, funded by MCIN/AEI/10.13039/501100011033.
ER acknowledges the support of the Natural Sciences and Engineering Research Council of Canada (NSERC), funding reference number RGPIN-2017-03987.
The work of JS is partially supported by the Natural Sciences and Engineering Research Council of Canada (NSERC) through the Canadian Institute for Theoretical Astrophysics (CITA) National Fellowship. 
ES and TGW acknowledge funding from the European Research Council (ERC) under the European Union’s Horizon 2020 research and innovation programme (grant agreement No. 694343). 
ATB and FB would like to acknowledge funding from the European Research Council (ERC) under the European Union’s Horizon 2020 research and innovation programme (grant agreement No.726384/Empire).
SCOG and RSK acknowledge financial support from the DFG via the Collaborative Research Center (SFB 881, Project-ID 138713538) `The Milky Way System' (subprojects A1, B1, B2, B8). They also acknowledge funding from the Heidelberg cluster of excellence (EXC 2181 - 390900948) `STRUCTURES: A unifying approach to emergent phenomena in the physical world, mathematics, and complex data', and from the European Research Council in the ERC synergy grant `ECOGAL – Understanding our Galactic ecosystem: From the disk of the Milky Way to the formation sites of stars and planets' (project ID 855130).
This paper makes use of the following ALMA data: \linebreak
ADS/JAO.ALMA\#2012.1.00650.S, \linebreak 
ADS/JAO.ALMA\#2013.1.01161.S, \linebreak 
ADS/JAO.ALMA\#2015.1.00956.S, \linebreak 
ADS/JAO.ALMA\#2017.1.00886.L, \linebreak 
ADS/JAO.ALMA\#2018.1.01651.S. \linebreak 
ALMA is a partnership of ESO (representing its member states), NSF (USA) and NINS (Japan), together with NRC (Canada), MOST and ASIAA (Taiwan), and KASI (Republic of Korea), in cooperation with the Republic of Chile. The Joint ALMA Observatory is operated by ESO, AUI/NRAO and NAOJ.
This paper includes data gathered with the 2.5 meter du Pont located at Las Campanas Observatory, Chile, and data based on observations carried out at the MPG 2.2m telescope on La Silla, Chile.

\section*{Data availability}
The data underlying this article will be shared on reasonable request to the corresponding author.

\bibliographystyle{mnras}
\bibliography{mybib}
 
\appendix
\section{Accuracy of our results}\label{sec:accuracy}
In order to validate the accuracy of our measurements, we verify that the requirements listed in sect. 4.4 of \citet{kruijssen18} are fulfilled. Satisfaction of these criteria indicates that the constrained parameters ($t_{\rm CO}$, $t_{\rm fb}$, and $\lambda$) are measured with an accuracy of at least 30\%. 

\begin{enumerate}
\item The duration of $t_{\rm CO}$ and $t_{\rm H\alpha}$ should differ by less than one order of magnitude. This is satisfied by $|\log_{10}(t_{\rm H\alpha}/t_{\rm CO})| \leq 0.58$. 
\item The ratio $\lambda/l_{\rm ap,min}$ ranges from 1.06 to 3.63 for the galaxies in our sample. For 8 galaxies out of 54 (IC1954, NGC1087, NGC1385, NGC1546, NGC1672,
NGC4298, NGC4540, NGC4548), we measure $\lambda/l_{\rm ap,min} \rm <1.5$, implying that for these galaxies, only $t_{\rm CO}$ is constrained with sufficient accuracy, while the constrained $t_{\rm fb}$ and $\lambda$ are upper limits. For the remaining galaxies, we ensure that the mean separation length between independent regions are sufficiently resolved by our observations. 

\item We ensure that the number of identified peaks in both CO and H$\rm\alpha$ emission maps is always above 35. Galaxies without enough peaks were initially removed from our galaxy sample as described in Section~\ref{sec:data}.
\item The CO-to-H$\rm\alpha$ flux ratios measured locally focusing on CO (H$\rm\alpha$) peaks should never be below (above) the galactic average. As shown in Figure~\ref{fig:tuningforks}, this criterion is fulfilled, implying that we filter out the large-scale diffuse emission that is not associated with peaks enclosed in the aperture. 
\item The global star formation history of the analysed region, during the last evolutionary cycle (ranging $9-35$\,Myr), should not vary more than 0.2\,dex, when averaged over time width of $t_{\rm CO}$ or $t_{\rm H\alpha}$. This is to ensure that we homogeneously sample the evolutionary timelines from gas to star with the identified peaks. Unfortunately, SFR over the last course of cloud lifetime for the galaxies in our sample are not known. However, we expect that the variation of SFR in the last $\sim 35$\,Myr to be minor when time averaged by $t_{\rm CO}$ or $t_{\rm H\alpha}$ as these low redshift galaxies are mostly undergoing a secular evolution, especially when the galactic centres and bars are excluded.
\item Each region, independently undergoing evolution from gas to star, should be detectable in both tracers at some point in their life. This implies that sensitivity of the CO and H$\rm\alpha$ should be matched, allowing the faintest CO peak to evolve into H\,{\sc ii} regions that is bright enough to be detected in the H$\rm\alpha$ map. In order to check if this criterion is satisfied, we first calculate the minimum star-forming region mass expected to form from the detected molecular clouds by multiplying the typical 5$\sigma$ point source sensitivity of the CO map ($\sim10^5\rm\,M_{\odot}pc^{-2}$; \citealp{leroy21_survey}) by the typical star formation efficiency constrained in our method ($\epsilon_{\rm sf}=2.8\pm 1.5\%$). We then compare this minimum mass to the mass of the stellar population required to produce ionising radiation that matches the typical 5$\sigma$ sensitivity of H$\rm\alpha$ map on the scale of the typical individual star-forming regions ($\lambda\approx250$\,pc). We use the \textsc{Starburst99} model \citep{leitherer99} to obtain the initial mass of the stellar population assuming that stars formation took place instantaneously 5\,Myr ago. We find that the typical minimum mass of the stellar population obtained from CO maps ($3000\, M_{\rm\odot}$) matches well with that from H$\rm\alpha$ maps ($4000\,M_{\rm\odot}$). 
\end{enumerate}

Our measurements satisfy almost all of the requirements listed above with an exception of (ii). This implies that while $t_{\rm CO}$ is constrained with high accuracy, we do not have sufficient resolution to precisely constrain the $\lambda$ and $t_{\rm fb}$, for 8 galaxies in our sample. Only upper limits can be obtained for these values. Below, we use four more criteria listed in \citet{kruijssen18} to further determine the validity of $t_{\rm fb}$. To do so, we first introduce the filling factor of SFR or gas tracer peaks, which is defined as $\zeta = 2r/\lambda$, where $r$ is the mean radius of the corresponding peaks. This $\zeta$ characterises how densely the peaks are located in a map. The average $\zeta$ is calculated by weighting the filling factors of gas and SFR tracer peaks with their associated time-scales. 

\begin{figure}
\includegraphics[width=\linewidth]{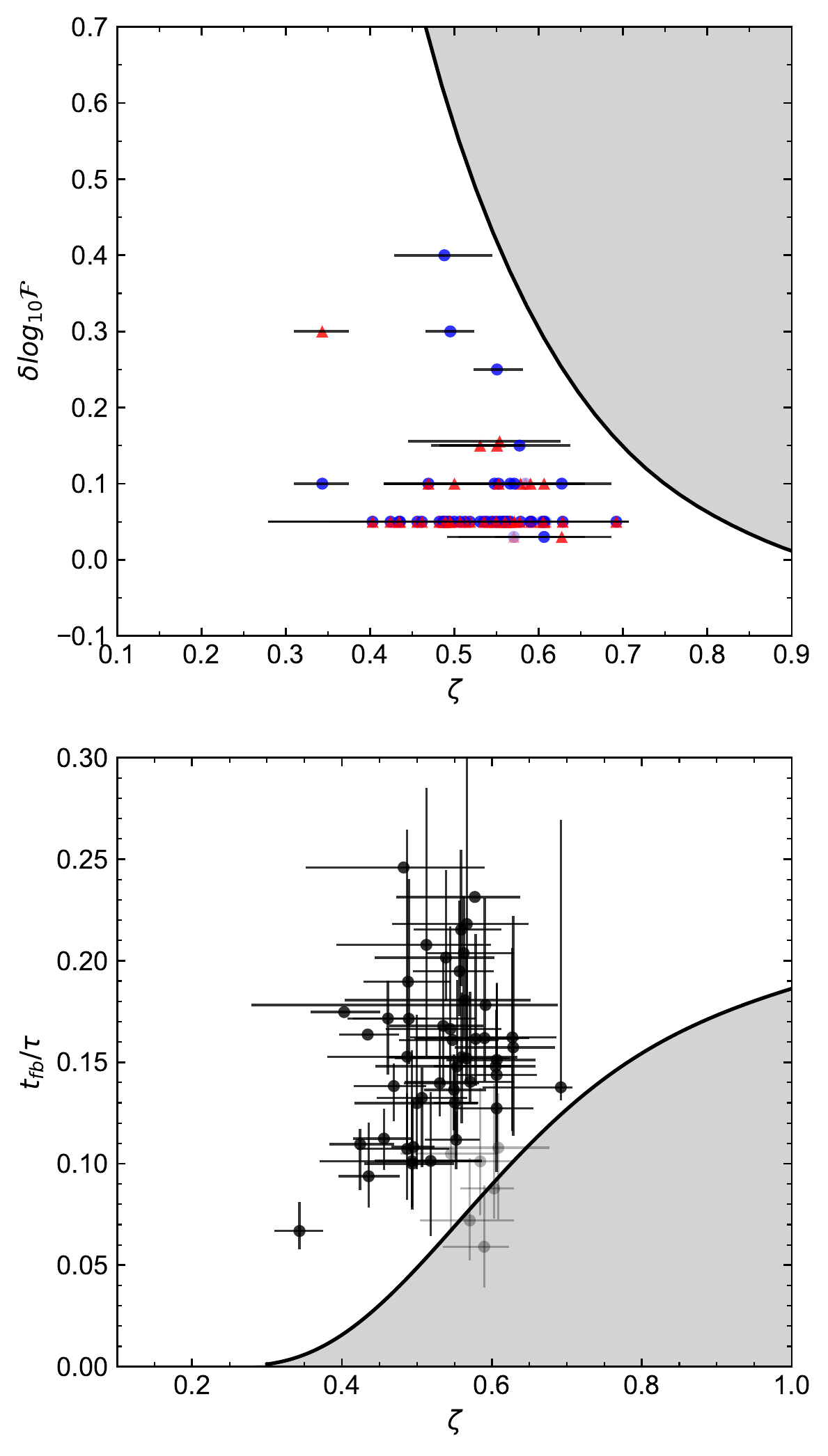}
\caption{In the top panel, we show the adopted density contrasts ($\delta\rm log_{10}\mathcal{F}$) used for the peak identification in each H$\rm\alpha$ (blue) and CO (red) emission map, as a function of the average filling factor $\zeta$. The shaded area is the parameter space where the peak identification is affected by blending of sources \citep{kruijssen18}. We confirm that we adopt small enough $\delta\rm log_{10}\mathcal{F}$ to identify adjacent peaks even in maps with high filling factor. In the bottom panel, we show the ratio of the feedback time-scale ($t_{\rm fb}$) and the total duration of the evolutionary cycle ($\tau$), as a function of the average filling factor. The shaded area is the parameter space where the contamination by adjacent peaks affects the measurement of the feedback time-scale. As a result, in gray we show six galaxies where only an upper limit of $t_{\rm fb}$ can be determined by not satisfying the condition (viii).}
\label{fig:blending}
\end{figure}

\begin{enumerate}\addtocounter{enumi}{6}
\item When peaks are densely distributed potentially overlapping with each other, the density contrast used for identifying peaks ($\delta\rm log_{10}\mathcal{F}$) should be small enough to identify adjacent peaks. In Figure~\ref{fig:blending}, we confirm that our adopted $\delta\rm log_{10}\mathcal{F}$ is small enough, compared to the upper limit prescribed by \citet{kruijssen18}.
\item Spatial overlap of adjacent peaks due to high filling factor can falsely be attributed to a longer duration of the measured feedback time-scale. In this case, only an upper limit on the feedback time-scale can be determined. In order to check whether we sufficiently resolve independent regions, we compare in Figure~\ref{fig:blending} the analytical prescription of \citet{kruijssen18} with our measurements of $t_{\rm fb}/\tau$, where $\tau$ is the total duration of the entire evolutionary cycle ($\tau=t_{\rm CO}+t_{\rm H\alpha}-t_{\rm fb}$). We find that this condition is not fulfilled for six galaxies (NGC1097, NGC2090, NGC3627, NGC4540, NGC4548, NGC4569), two of which overlap with galaxies that do not satisfy condition (ii). 
\item As shown in the lower panel of Figure~\ref{fig:blending}, we ensure the conditions $t_{\rm fb} > 0.05 \tau$ and $t_{\rm fb} < 0.95 \tau$ are verified for all galaxies.
\item Similarly to condition (v), the global SFR of the analysed region should not vary more than 0.2\,dex during the entire evolutionary lifecycle when averaged over $t_{\rm fb}$. This is satisfied using the same argument in (v) stated above.
\item After masking obviously crowed regions such as the galaxy centre, visual inspection does not reveal abundant region blending. 

In conclusion, we find that most of our measurements are constrained with high accuracy. The only exceptions are $\lambda$ in 8 galaxies that do not satisfying condition (ii) and $t_{\rm fb}$ in 12 galaxies that do not satisfy both conditions (ii) and (viii). Only upper limits for these values can be constrained for this subset of galaxies.

\end{enumerate}

\vspace{4mm}
\noindent {\it
$^1$Astronomisches Rechen-Institut, Zentrum f\"{u}r Astronomie der Universit\"{a}t Heidelberg, M\"{o}nchhofstra\ss e 12-14, 69120 Heidelberg, Germany\\ 
$^{2}$Instit\"ut f\"{u}r Theoretische Astrophysik, Zentrum f\"{u}r Astronomie der Universit\"{a}t Heidelberg, Albert-Ueberle-Strasse 2, 69120 Heidelberg, Germany\\ 
$^{3}$Department of Astronomy, The Ohio State University, 140 West 18th Ave, Columbus, OH 43210, USA\\ 
$^4$Max-Planck Institut f\"ur Extraterrestrische Physik, Giessenbachstra\ss e 1, 85748 Garching, Germany\\ 
$^{5}$Argelander-Institut f\"{u}r Astronomie, Universit\"{a}t Bonn, Auf dem H\"{u}gel 71, 53121 Bonn, Germany\\ 
$^{6}$The Observatories of the Carnegie Institution for Science, 813 Santa Barbara Street, Pasadena, CA 91101, USA\\ 
$^{7}$Departamento de Astronomía, Universidad de Chile, Casilla 36-D, Santiago, Chile\\ 
$^8$Department of Physics \& Astronomy, University of Wyoming, Laramie, WY 82071\\
$^{9}$Department of Physics, University of Connecticut, 196A Auditorium Road, Storrs, CT 06269\\
$^{10}$Research School of Astronomy and Astrophysics, Australian National University, Canberra, ACT 2611, Australia\\ 
$^{11}$ARC Centre of Excellence for All Sky Astrophysics in 3 Dimensions (ASTRO 3D), Australia\\
$^{12}$International Centre for Radio Astronomy Research, University of Western Australia, 7 Fairway, Crawley, 6009, WA, Australia\\ 
$^{13}$CNRS, IRAP, 9 Av. du Colonel Roche, BP 44346, F-31028 Toulouse cedex 4, France\\  $^{14}$Universit\'{e} de Toulouse, UPS-OMP, IRAP, F-31028 Toulouse cedex 4, France\\
$^{15}$Universit\"{a}t Heidelberg, Interdisziplin\"{a}res Zentrum f\"{u}r Wissenschaftliches Rechnen, Im Neuenheimer Feld 205, 69120 Heidelberg, Germany\\ 
$^{16}$School of Mathematics and Physics, University of Queensland, St Lucia 4067, Australia\\
$^{17}$Department of Physics, Tamkang University, No.151, Yingzhuan Road, Tamsui District, New Taipei City 251301, Taiwan\\
$^{18}$IRAM, 300 rue de la Piscine, 38406 Saint Martin d'H\`eres, France\\
$^{19}$Sorbonne Universit\'e, Observatoire de Paris, Universit\'e PSL, CNRS, LERMA, F-75005, Paris, France\\
$^{20}$Observatorio Astron{\'o}mico Nacional (IGN), C/Alfonso XII 3, Madrid E-28014, Spain\\ 
$^{21}$4-183 CCIS, University of Alberta, Edmonton, Alberta, Canada\\
$^{22}$National Astronomical Observatory of Japan, 2-21-1 Osawa, Mitaka, Tokyo, 181-8588, Japan\\
$^{23}$Max Planck Institut für Astronomie, Königstuhl 17, 69117 Heidelberg, Germany\\
$^{24}$Department of Physics and Astronomy, McMaster University, 1280 Main Street West, Hamilton, ON L8S 4M1, Canada\\
$^{25}$Canadian Institute for Theoretical Astrophysics (CITA), University of Toronto, 60 St George Street, Toronto, ON M5S 3H8, Canada\\
$^{26}$Dipartimento di Fisica e Astronomia, Università di Firenze, Via G. Sansone 1, 50019 Sesto Fiorentino, Firenze, Italy\\
$^{27}$INAF – Osservatorio Astrofisico di Arcetri, Largo E. Fermi 5, 50127 Firenze, Italy\\
$^{28}$Observatorio Astron{\'o}mico Nacional (IGN), C/Alfonso XII 3, Madrid E-28014, Spain\\ 
}

\bsp
\label{lastpage}
\end{document}